\documentclass[aip,preprint]{revtex4-2}
\usepackage[cp1251]{inputenc}
\usepackage[T2A]{fontenc}
\usepackage[english]{babel}
\usepackage{amssymb,latexsym,amsmath,amscd}
\usepackage{graphicx,color,framed}
\usepackage{xcolor}
\usepackage{siunitx}
\usepackage{empheq}
\usepackage{xr}
\usepackage{microtype}
\usepackage{amsfonts}
\usepackage[normalem]{ulem}
\usepackage{caption}
\usepackage{subcaption}

\begin{document}
\title{Theory of electrolyte solutions in a slit charged pore: effects of structural interactions and specific adsorption of ions}
%\thanks{}
\author{\firstname{Victoria A.} \surname{Vasileva}}
\affiliation{School of Applied Mathematics, HSE University, Tallinskaya st. 34, 123458 Moscow, Russia}
\author{\firstname{Daria A.} \surname{Mazur}}
\affiliation{School of Applied Mathematics, HSE University, Tallinskaya st. 34, 123458 Moscow, Russia}
\author{\firstname{Yury A.} \surname{Budkov}}
\email[]{ybudkov@hse.ru}
\affiliation{School of Applied Mathematics, HSE University, Tallinskaya st. 34, 123458 Moscow, Russia}
\affiliation{Laboratory of Computational Physics, HSE University, Tallinskaya st. 34, 123458 Moscow, Russia}
\affiliation{Laboratory of Multiscale Modeling of Molecular Systems,G.A. Krestov Institute of Solution Chemistry of the Russian Academy of Sciences, 153045, Akademicheskaya st. 1, Ivanovo, Russia}
\begin{abstract}
In this paper, we present a continuation of our research on modeling electrolyte solutions within charged slit pores. We make use of the model developed by Blossey {\sl et al.}, which takes into account the structural interactions between ions through a bilinear form over the gradients of local ionic concentrations in the grand thermodynamic potential, as well as their steric interactions through the lattice gas model. The structural interactions may describe effects of the molecular structure of ions at a phenomenological level. For example, these effects include steric effects due to non-spherical shapes of ions, their conformation lability, and solvent effects. In addition, we explore their specific interactions with the pore walls by incorporating external attractive potentials. Our primary focus is on observing the behavior of ionic concentration profiles and the disjoining pressure as the pore width changes. By starting with the local mechanical equilibrium condition, we derive a general expression for the disjoining pressure. Our findings indicate that considering the structural interactions of ions leads to a pronounced minimum on the disjoining pressure profiles at small pore widths. We attribute this minimum to the formation of electric double layers on the electrified surfaces of the pore. Additionally, our results demonstrate that inclusion of the attractive interactions of ions with the pore walls enhances this minimum and shifts it to smaller pore thicknesses. Our theoretical discoveries may be useful for those involved in supercapacitor electrochemical engineering, particularly when working with porous electrodes that have been infused with concentrated electrolyte solutions.
\end{abstract}
\maketitle

\section{Introduction}
The study of electrolyte solutions (ES) in confined geometries, such as charged pores or slit-like channels, has gained significant importance due to its inherent involvement in various scientific and technological applications, ranging from energy storage devices~\cite{hassan2023effect,xia2017electrolytes,zheng2017research}, to water purification systems~\cite{xu2023electrochemical,sharma2022biomimetic,aluru2023fluids,shin2017hydrocarbon}, and even to biological systems, including ion transport in cellular and ion-exchanged membranes~\cite{luo2018selectivity}. One of the critical challenges in understanding these systems is the accurate modeling of structural interactions of ions within the confinement, as well as the coupling between ion-specific effects and the electrostatic interactions~\cite{ben2011dielectric,frydel2011polarizable,hatlo2012electric,goodwin2017mean,uematsu2018effects,budkov2021theory,budkov2020two,podgornik}. To address this challenge, the self-consistent field theory (SCFT) has emerged as a powerful and versatile tool in modeling the behavior of ionic liquid-phase systems~\cite{naji2013perspective,blossey2023poisson,budkov2021electric}.

To effectively study electrolyte solutions confined within solid nanostructures of varying shapes (e.g. nano-sized pores), it is essential to calculate the mechanical stress using the stress tensor, along with ionic concentrations and electrostatic potential profiles. Calculating the local stress tensor, corresponding to certain SCF equations, we can determine practically important physical properties like solvation pressure and shear stresses~\cite{misra2019theory,de2020continuum,de2020interfacial,shi2023perspective,kolesnikov2021models,gurina2022disjoining,kolesnikov2022electrosorption}. Such properties can help estimate the porous material deformation of a particular elastic modulus, which is crucial for batteries and supercapacitors that use microporous electrodes impregnated with liquid-phase electrolytes (see, for instance, ref. \cite{koczwara2017situ,chen2022porous,da2020reviewing,li2018high,gurina2022disjoining,kolesnikov2022electrosorption}). Additionally, the stress tensor can be used to measure the macroscopic force exerted on charged macroscopic conductors or dielectrics that are immersed in ionic liquids. As such, a first-principles approach that enables us to extract the stress tensor of inhomogeneous ionic liquids from the thermodynamic potential would be valuable for practical purposes.

Some progress have been recently achieved in this area~\cite{budkov2022modified,budkov2023macroscopic,brandyshev2023noether}. 
In ref.~\cite{budkov2022modified}, Budkov and Kolesnikov attempted to apply Noether's theorem to the grand thermodynamic potential of an ionic liquid as a functional of the electrostatic potential. The authors established a conservation law, $\partial_{i}\sigma_{ik}=0$, which represents the local mechanical equilibrium condition in terms of the symmetric stress tensor, $\sigma_{ik}$. The stress tensor obtained consisted of two components: the electrostatic (Maxwell) stress tensor and the hydrostatic isotropic stress tensor. The former is associated with the local electric field, while the latter, in general case~\cite{budkov2022modified,gongadze2014ions,marcovitz2015water}, is determined by both the local field and potential. The authors generalized the local mechanical equilibrium condition for the cases when external potential forces act on the ions. They then derived a general analytical expression for the electrostatic disjoining pressure of an ionic liquid confined in a charged nanopore slit, which extended the well-known DLVO expression to different reference models of liquid. Budkov and Kalikin presented a SCFT of macroscopic forces in inhomogeneous flexible chain polyelectrolyte solutions in ref.\cite{budkov2023macroscopic}. The authors derived an analytical expression for a stress tensor by subjecting the system to a small dilation and considering self-consistent field equations resulting from the extremum condition of the grand thermodynamic potential. This stress tensor, in addition to the previously mentioned hydrostatic and Maxwell stress tensors, also includes a conformational stress tensor generated by the conformational (Lifshitz) entropy of flexible polymer chains. The authors applied their theory to the investigation of polyelectrolyte solution constrained in a conducting slit nanopore and noted anomalies in disjoining pressure and electric differential capacitance at small pore thicknesses. Brandyshev and Budkov~\cite{brandyshev2023noether} proposed the general covariant approach based on Noether's second theorem allowing them to derive the symmetric stress tensor from a grand thermodynamic potential for an arbitrary model of inhomogeneous liquid. They applied their approach to several models of inhomogeneous ionic liquids that consider electrostatic correlations of ions or short-range correlations related to packing effects. Specifically, they derived analytical expressions for the symmetric stress tensors of the Cahn-Hilliard-like model~\cite{maggs2016general}, Bazant-Storey-Kornyshev model~\cite{bazant2011double}, and Maggs-Podgornik-Blossey model~\cite{maggs2016general}.

In this paper, we model liquid electrolytes in charged pores, utilizing the Cahn-Hilliard-like model~\cite{blossey2017structural,blossey2023poisson}. This model considers the structural and steric interactions in the grand thermodynamic potential. Theory also takes into account the short-range specific interactions between the ions and the pore walls through attractive external potentials (specific adsorption). Within this approach, we investigate the behavior of the disjoining pressure in connection with the local ion concentrations.

\section{Theoretical background}
We consider an electrolyte solution model with account for the so-called structural interactions~\cite{blossey2017structural} within the Cahn-Hilliard~\cite{cahn1965phase} approach, which takes into account molecular structure effects via the bilinear form over the concentration gradients~\cite{cahn1965phase,blossey2017structural}. Thus, the grand thermodynamic potential (GTP) of the ES has the form 
\begin{equation}
\Omega = F_{el}+\Omega_{liq},
\end{equation}
where
\begin{equation}
F_{el}=\int d\bold{r}\left(-\frac{\epsilon \epsilon_0 \left(\nabla\psi\right)^2}{2}+\rho\psi\right)
\end{equation}
is the electrostatic free energy of the ES with the local electrostatic potential $\psi(\bold{r})$; the local charge density of ions is $\rho(\bold{r})=\sum_{\alpha}q_{\alpha}c_{\alpha}(\bold{r})$, $q_{\alpha}$ is the electrostatic charge of ion of the $\alpha$th type; $\epsilon$ is the permittivity of solvent; we will model the latter as a uniform dielectric medium; $\epsilon_0$ is the vacuum permittivity; $c_{\alpha}(\bold{r})$ are the local ionic concentrations. The GTP of the reference liquid system is
\begin{equation}
\label{omega_liquid}
\Omega_{liq} = \int d\bold{r}\left(f(\{c_{\alpha}\})+\sum\limits_{\alpha}c_{\alpha}w_{\alpha}+\frac{1}{2}\sum\limits_{\alpha\gamma}\varkappa_{\alpha\gamma}\nabla c_{\alpha}\cdot\nabla c_{\gamma} -\sum\limits_{\alpha}\mu_{\alpha}c_{\alpha}\right),
\end{equation}
where the first term in the integrand is the free energy density of liquid as the function of the local ionic concentrations, $c_{\alpha}$; the second term is the potential energy density of ions in the external potential fields with potential energies $w_{\alpha}(\bold{r})$; the third term is the contribution of the so-called structural interactions of the ions within the Cahn-Hilliard approach with the structural constants $\varkappa_{\alpha\gamma}$ (see ref.~\cite{blossey2017structural}), $\mu_{\alpha}$ are the bulk ionic chemical potentials. Note that in the general case, the structural interactions may describe effects related to the molecular structure of ions at a phenomenological level. Some examples of these effects include steric effects due to non-spherical shapes of ions, their conformation lability, and solvent effects. We do not specify here the physical nature of the structural contribution to the GTP, as well as the nature of the external potentials, wondering how it influences the local concentration profiles and disjoining pressure (see below) relative to the regular modified Poisson-Boltzmann theory~\cite{borukhov1997steric,kornyshev2007double,kralj1996simple}.

Thus, the GTP can be rewritten as follows
\begin{equation}
\label{omega_liquid_2}
\Omega =\int d\bold{r}\omega(\bold{r}),
\end{equation}
where the GTP density is
\begin{equation}
\label{omega}
\omega=-\frac{\epsilon \epsilon_0\left(\nabla\psi\right)^2}{2}+\rho\psi+\frac{1}{2}\sum\limits_{\alpha\gamma}\varkappa_{\alpha\gamma}\nabla c_{\alpha}\cdot\nabla c_{\gamma}+f(\{c_{\alpha}\})+\sum\limits_{\alpha}c_{\alpha}w_{\alpha}-\sum\limits_{\alpha}\mu_{\alpha}c_{\alpha}.
\end{equation}
Note that summations in (\ref{omega}) are performed over all kinds of the ions in the ES. Note also that integration in (\ref{omega_liquid_2}) is performed over the volume of the ES bounded by the surfaces of conducting or dielectric macroscopic bodies. The Euler-Lagrange equations, 
\begin{equation}
\frac{\delta\Omega}{\delta\psi}=\frac{\partial \omega}{\partial{\psi}}-\partial_{i}\left(\frac{\partial\omega}{\partial\psi_{,i}}\right)=0,~\frac{\delta\Omega}{\delta c_{\alpha}}=\frac{\partial \omega}{\partial{c_{\alpha}}}-\partial_{i}\left(\frac{\partial\omega}{\partial c_{\alpha,i}}\right)=0,
\end{equation}
yield
\begin{equation}
\label{EL_CH}
\nabla^2\psi = -\frac{1}{\epsilon\epsilon_0}\sum\limits_{\alpha}q_{\alpha}c_{\alpha},~\sum\limits_{\gamma}\varkappa_{\alpha\gamma}\nabla^2 c_{\gamma} = v_{\alpha}, 
\end{equation}
where $v_{\alpha}=q_{\alpha}\psi+w_{\alpha}+\bar{\mu}_{\alpha}-\mu_{\alpha}$, where $\bar{\mu}_{\alpha}=\partial{f}/\partial{c_{\alpha}}$ are the intrinsic chemical potential of the ions in the local density approximation. The first equation is the standard Poisson equation for the electrostatic potential, whereas the second one describes the thermodynamic equilibrium condition for ions of the $\alpha^{th}$ kind. The system of these self-consistent field equations can be solved with the appropriate boundary conditions for the electrostatic potential and local ionic concentrations. Note that the boundary conditions for the Poisson equation are determined by the nature of the macroscopic bodies immersed in the ionic liquid. In what follows, we will consider the case of conducting body with fixed surface potential, immersed into the ES. For the local ionic concentrations we assume that at the surface of the immersed bodies $c_{\alpha}=0$. The latter boundary condition is related to the fact that the ions undergo strong repulsive forces in close proximity to the surface. We assume that the specific attractive interactions of the ions with the immersed bodies (specific adsorption) is included in the potential energies $w_{\alpha}(\bold{r})$. Note that the short-range specific interactions between ions~\cite{goodwin2017mean,budkov2018theory} are beyond the scope of this work. We also do not explicitly consider the polar or polarizable solvent molecules~\cite{iglivc2010excluded,budkov2018theory,budkov2022modified,budkov2016theory,budkov2015modified}. Furthermore, we neglect the orientation and static polarizability of the ions~\cite{hatlo2012electric,frydel2011polarizable,budkov2020two,budkov2021theory,budkov2022modified}. We do not discuss the effect of dielectric decrement~\cite{ben2011dielectric,nakayama2015differential} and electrostatic correlations~\cite{bazant2011double,de2020continuum} either. Although these effects can be directly incorporated into the current theory, they are irrelevant to the physical effects that are being discussed below.

Then, with using eq. (\ref{EL_CH}), we get
\begin{equation}
\label{Noether}
\frac{\partial\omega}{\partial x_{i}}=\frac{\partial}{\partial{x_{k}}}\left(\psi_{,i}\frac{\partial{\omega}}{\partial{\psi_{,k}}}\right)+\frac{\partial}{\partial x_{k}}\left(\sum\limits_{\alpha}c_{\alpha,i}\frac{\partial{\omega}}{\partial c_{\alpha,k}}\right)+\sum\limits_{\alpha}c_{\alpha}\frac{\partial{w}_{\alpha}}{\partial{x}_{i}}.
\end{equation}
or 
\begin{equation}
\label{equil}
\frac{\partial \sigma_{ik}}{\partial{x_{k}}}-\sum\limits_{\alpha}c_{\alpha}\frac{\partial w_{\alpha}}{\partial x_{i}}=0,
\end{equation}
where $c_{\alpha,i}=\partial_{i} c_{\alpha}$, $\psi_{,i}=\partial_{i}{\psi}$, $\partial_{i}=\partial/\partial{x_i}$ and
\begin{equation}
\sigma_{ik}=\omega \delta_{ik}-\psi_{,i}\frac{\partial{\omega}}{\partial{\psi_{,k}}}-\sum\limits_{\alpha}c_{\alpha,i}\frac{\partial{\omega}}{\partial c_{\alpha,k}}
\end{equation}
is the total stress tensor~\cite{brandyshev2023noether,budkov2022modified,budkov2023macroscopic}. Note that a summation over the repeated coordinate indices in (\ref{Noether}) and (\ref{equil}) is implied. With the use of the SCF equations (\ref{EL_CH}), the total stress tensor can be divided into three terms
\begin{equation}
\label{stress}
\sigma_{ik}=\sigma_{ik}^{(M)}+\sigma_{ik}^{(h)}+\sigma_{ik}^{(s)},
\end{equation}
where
\begin{equation}
\label{Maxwell}
\sigma_{ik}^{(M)}=\epsilon \epsilon_0\bigg(E_{i}E_{k}-\frac{1}{2}\bold{E}^{2}\delta_{ik}\bigg) 
\end{equation}
is the Maxwell electrostatic stress tensor with the electric field components, $E_{i}=-\partial_{i}{\psi}$,
\begin{equation}
\label{hyd}
\sigma_{ik}^{(h)}=-P\delta_{ik}    
\end{equation}
is the hydrostatic stress tensor with the local osmotic pressure, $P=\sum_{\alpha}c_{\alpha}\partial{f}/\partial{c_{\alpha}}-f$, of the ions, whereas
\begin{equation}
\label{stress_struct}
\sigma_{ik}^{(s)}=\sum\limits_{\alpha\gamma}\varkappa_{\alpha\gamma}\left[\left(\frac{1}{2}\nabla c_{\alpha}\cdot \nabla c_{\gamma} +c_{\alpha}\nabla^2c_{\gamma}\right)\delta_{ik}-\partial_{i}c_{\alpha}\partial_{k}c_{\gamma}\right]
\end{equation}
is the contribution of the structural interactions to the total stress tensor. Note that the stress tensor (\ref{stress_struct}) has been recently obtained by Brandyshev and Budkov within a general covariant approach~\cite{brandyshev2023noether}.

Eq. (\ref{equil}) is nothing but a local mechanical equilibrium condition of the ES which will be used in the next section to derive the expression for the disjoining pressure of the ES in a slit-like pore.

\section{Disjoining pressure}
Now we would like to discuss how we can derive the general mean-field approximation of the disjoining pressure for an ES in a slit charged nanopore. Placing the origin of $z$ axis on one charged wall and another one at $z=H$, we can calculate the disjoining pressure by the standard relation~\cite{kolesnikov2021models}
\begin{equation}
\Pi=-\frac{\partial(\Omega /\mathcal{A})}{\partial H}-P_b,
\end{equation}
where $P_{b}$ is the ES osmotic pressure in the bulk and 
\begin{equation}
\Omega/\mathcal{A} = \int\limits_{0}^{H}\omega(z)dz
\end{equation}
is the GTP per unit area of the pore walls, $\mathcal{A}$ is the total area of the pore walls. Thus, we have
\begin{equation}
\nonumber
\frac{1}{\mathcal{A}}\frac{\partial\Omega}{\partial H}=\int\limits_{0}^{H} dz\left(\frac{\delta\Omega}{\delta\psi(z)}\frac{\partial\psi(z)}{\partial H}+\sum\limits_{\alpha}\frac{\delta\Omega}{\delta c_{\alpha}(z)}\frac{\partial c_{\alpha}(z)}{\partial H}\right)+\sigma_{zz}(H)+\int\limits_{0}^{H}dz\sum\limits_{\alpha}\frac{\partial \omega}{\partial w_{\alpha}}\frac{\partial w_{\alpha}}{\partial H}
\end{equation}
\begin{equation}
= \int\limits_{0}^{H}dz\sum\limits_{\alpha}{c}_{\alpha}(z)\frac{\partial w_{\alpha}(z)}{\partial H}+\sigma_{zz}(H),
\end{equation}
where we took into account that ${\partial \omega}/{\partial w_{\alpha}}=c_{\alpha}$ and ${\delta\Omega}/{\delta\psi(z)}=0$, ${\delta\Omega}/{\delta c_{\alpha}(z)}=0$. Therefore, we eventually obtain
\begin{equation}
\Pi=-\sum\limits_{\alpha}\int\limits_{0}^{H}dz c_{\alpha}(z)\frac{\partial w_{\alpha}(z)}{\partial H}-\sigma_{zz}(H)-P_b,
\end{equation}
where $\sigma_{zz}(H)=(\omega-\psi^{\prime}{\partial\omega}/{\partial\psi^{\prime}}-\sum_{\alpha}c_{\alpha}^{\prime}{\partial{\omega}}/{\partial c_{\alpha}^{\prime}})|_{z=H}$ is the normal stress at $z=H$.

Now let us consider the special practically important case, when the external potentials are created by identical walls, i.e. 
\begin{equation}
\label{u}
w_{\alpha}(z)=u_{\alpha}(z)+u_{\alpha}(H-z),
\end{equation}
where $u_{\alpha}$ is the single wall potential. Then, taking into account that ${c}_{\alpha}(z)={c}_{\alpha}(H-z)$, we arrive at
\begin{equation}
\label{Pi}
\Pi=-\sum\limits_{\alpha}\int\limits_{0}^{H}dz{c}_{\alpha}(z)u_{\alpha}^{\prime}(z)-\sigma_{zz}(H)-P_b.
\end{equation}
Eq. (\ref{Pi}) can be rewritten in a form that is more useful for practical calculations. Using the local mechanical equilibrium condition
\begin{equation}
\frac{d\sigma_{zz}(z)}{dz}-\sum\limits_{\alpha}{c}_{\alpha}(z)w_{\alpha}^{\prime}(z)=0,
\end{equation}
after the integration from $z=H/2$ to $z=H$ we obtain
\begin{equation}
\label{sigmazz}
\sigma_{zz}(H)=\sigma_{zz}\left(\frac{H}{2}\right)+\sum\limits_{\alpha}\int\limits_{H/2}^{H}dz{c}_{\alpha}(z)u_{\alpha}^{\prime}(z)-\sum\limits_{\alpha}\int\limits_{0}^{H/2}dz{c}_{\alpha}(z)u_{\alpha}^{\prime}(z),
\end{equation}
where we accounted for eq. (\ref{u}) and mentioned equality $c_{\alpha}(z)=c_{\alpha}(H-z)$.
Substituting expression (\ref{sigmazz}) for $\sigma_{zz}(H)$ in eq.(\ref{Pi}), after some algebra, we obtain
\begin{equation}
\label{Pi2}
\Pi=P_{n}-P_{b}-2\sum\limits_{\alpha}\int\limits_{H/2}^{H}dz c_{\alpha}(z)u_{\alpha}^{\prime}(z),
\end{equation}
where 
\begin{equation}
\label{normal_press}
P_{n}=-\sigma_{zz}\left(\frac{H}{2}\right)=P_m-\sum\limits_{\alpha\gamma}\varkappa_{\alpha\gamma}c_{\alpha}\left(\frac{H}{2}\right)c_{\gamma}^{\prime\prime}\left(\frac{H}{2}\right)
\end{equation}
is the normal osmotic pressure in the pore middle, where we took into account that $\psi^{\prime}(H/2)=c_{\alpha}^{\prime}(H/2)=0$ and introduced the osmotic pressure of the ions in the midpoint of the pore, $P_m=P\left(H/2\right)$. Using the SCF equations (\ref{EL_CH}), eq. (\ref{normal_press}) can be simplified to
\begin{equation}
\label{normal_press_2}
P_{n}=P_m-\sum\limits_{\alpha}c_{\alpha,m}v_{\alpha,m},
\end{equation}
where $c_{\alpha,m}=c_{\alpha}\left(H/2\right)$, $v_{\alpha,m}=v_{\alpha}\left(H/2\right)$. The second term in the right hand side of eq. (\ref{normal_press_2}) describes contribution of the structural interactions to the normal osmotic pressure of the ions in the slit pore with the identical charged walls.

The main analytical result of this work, eq. (\ref{Pi2}) alongside with eq.(\ref{normal_press_2}), is a generalization of the previously obtained disjoining pressure within the modified Poisson-Boltzmann theory~\cite{budkov2022modified} for the case of an account of the structural interactions of the ions. To calculate the disjoining pressure, it is necessary to solve the SCF equations (\ref{EL_CH}) with the appropriate boundary conditions.

\section{Numerical results and discussion}

\subsection{Basic model}
Before numerical calculations, we would like to specify the model of the ES confined in a slit electrified pore with identical walls possessing the fixed surface electrostatic potential, $\psi_0$. Following Maggs and Podgornik~\cite{maggs2016general}, we use the asymmetric lattice gas model which allows us to account for the ionic size asymmetry. The free energy density is~\cite{maggs2016general}
\begin{equation}
f = \frac{k_{B}T}{v}\left(\dfrac{\phi_+}{N_+} \ln \phi_+ + \dfrac{\phi_-}{N_-} \ln \phi_- +(1 - \phi_+ - \phi_-)\ln(1 - \phi_+ - \phi_-)\right),
\end{equation}
where $v$ is a cell volume of the lattice gas, $\phi_\pm = N_\pm c_\pm v$ are the local volume fractions of the ions, $N_{\alpha}$ -- number of cells occupied by ion of the $\alpha$th kind ($\alpha=\pm$); $k_B$ is the Boltzmann constant, $T$ is the temperature. The intrinsic chemical potentials are
\begin{equation}
    \bar \mu_\pm = \dfrac{\partial f}{\partial c_\pm} = {k_{B}T} \left( \ln{\phi_\pm} + 1 - N_{\pm}\ln\left(1 - \phi_+ - \phi_{-}\right) - N_{\pm}\right),
\end{equation}
and the bulk chemical potentials
\begin{equation}
\mu_{\pm} = k_{B}T\left(\ln\phi_\pm^{(b)} + 1 - N_{\pm}(\ln{(1 - \phi_+^{(b)} - \phi_-^{(b)})} + 1)\right),
\end{equation}
where $\phi^{(b)}_{\pm} = N_{\pm}  c^{(b)}_{\pm}v$ are the bulk volume fractions of the ions. Assuming that $q_\pm = \pm q$, we obtain from the bulk electroneutrality condition, $q_+ c^{(b)}_{+} + q_- c^{(b)}_{-} = 0$, that $c^{(b)}_\pm = c$. To describe interactions of the ions with pore wall, we use the following Gaussian well short-range potentials
\begin{equation}
\label{gaussian_well}
u_\pm(z) = -\varepsilon_{\pm} \exp\left[-\frac{z^2}{2 \sigma^2_\pm}\right],
\end{equation}
where $\varepsilon_{\pm}$ are the positive energetic parameters (depths) and $\sigma_{\pm}$ are the effective widths of the Gaussian wells.

The SCF equations and the boundary conditions in the dimensionless units take the form
\begin{equation}
\label{dimensionless_system}
\begin{cases}
\tilde c^{''}_\alpha (x) = \sum_{\gamma = \pm} \hat \chi_{\alpha \gamma} \tilde v_\gamma(x), \quad &\alpha = \pm \\
u^{''}(x) = \frac{1}{2}(\tilde c_-(x) - \tilde c_+(x)), &x \in [0, \tilde H] \\
\tilde c_{\alpha}(0) = \tilde c_{\alpha}(\tilde H) = 0, \\
u(0) = u(\tilde H) = u_0,
\end{cases}
\end{equation}
where we have introduced the following dimensionless variables: $\tilde c_\pm = {c_\pm}/{c}$, $c^{(b)}_\pm = c$, $u = {q \psi}/{k_B T}$, $u_0 = {q \psi_0}/{k_B T}$, $x ={z}/{r_D}$,  $\tilde{H} = H/{r_D}$, $\chi_{\alpha \gamma} = (k_B T r^2_D/c)^{-1}\varkappa _{\alpha \gamma}$, $\hat{\chi}_{\alpha\gamma}=\chi^{-1}_{\alpha\gamma}$, $\tilde v_\alpha = {v_\alpha}/{k_B T}$; $r_D = \sqrt{\epsilon \epsilon_0 k_B T/2 q^2 c}$ is the Debye length. The dimensionless disjoining pressure, $\tilde{\Pi}=\Pi/(ck_{B}T)$, can be written as follows
\begin{equation}
\begin{aligned}
\tilde \Pi = - \dfrac{1}{\phi_0} \cdot \ln\dfrac{1 - (N_+ \tilde c_{+ \, m} + N_-\tilde c_{- \, m})\phi_0}{1 - (N_+ + N_-)\phi_0}- N_+(\tilde c_{+ \, m} - 1)- N_-(\tilde c_{- \, m} - 1)& \\ + \tilde c_{+ \, m} + \tilde c_{- \, m} - 2 - \tilde c_{+ \, m} \tilde v_{+ \, m} - \tilde c_{- \, m} \tilde v_{- \, m}-2\sum\limits_{\alpha=\pm}\int\limits_{\tilde{H}/2}^{\tilde{H}}dx \tilde{c}_{\alpha}(x)\tilde{u}_{\alpha}^{\prime}(x),&
\end{aligned}
\end{equation}
where the parameter $\phi_0 = cv$ determines the packing of the ions in the bulk; the subscript "$m$" denotes that corresponding variables to be calculated at $x=\tilde{H}/2$; $\tilde{u}_{\alpha}=u_{\alpha}/k_{B}T$. For numerical calculations, we use the following physical parameters: $T=300$~K, $q=1.66\times 10^{-19}$~C, $\epsilon=40$, $c=1$~mol/l, $v^{1/3}=0.5$~nm. The Debye radius in this case is $r_D\approx 0.2$~nm and packing parameter $\phi_0\approx 0.08$. Thus, we consider the case of rather concentrated 1:1 ES with organic polar solvent like acetonitrile.

After numerical solution of the SCF equations (\ref{dimensionless_system}), we return to the physical units, in accordance with the above definitions. Numerical calculations were performed using the $solve\_bvp$ method from the Scipy library in Python\cite{virtanen2020scipy}.

\begin{figure}[ht]
  \centering
  \includegraphics[width=0.8\linewidth]{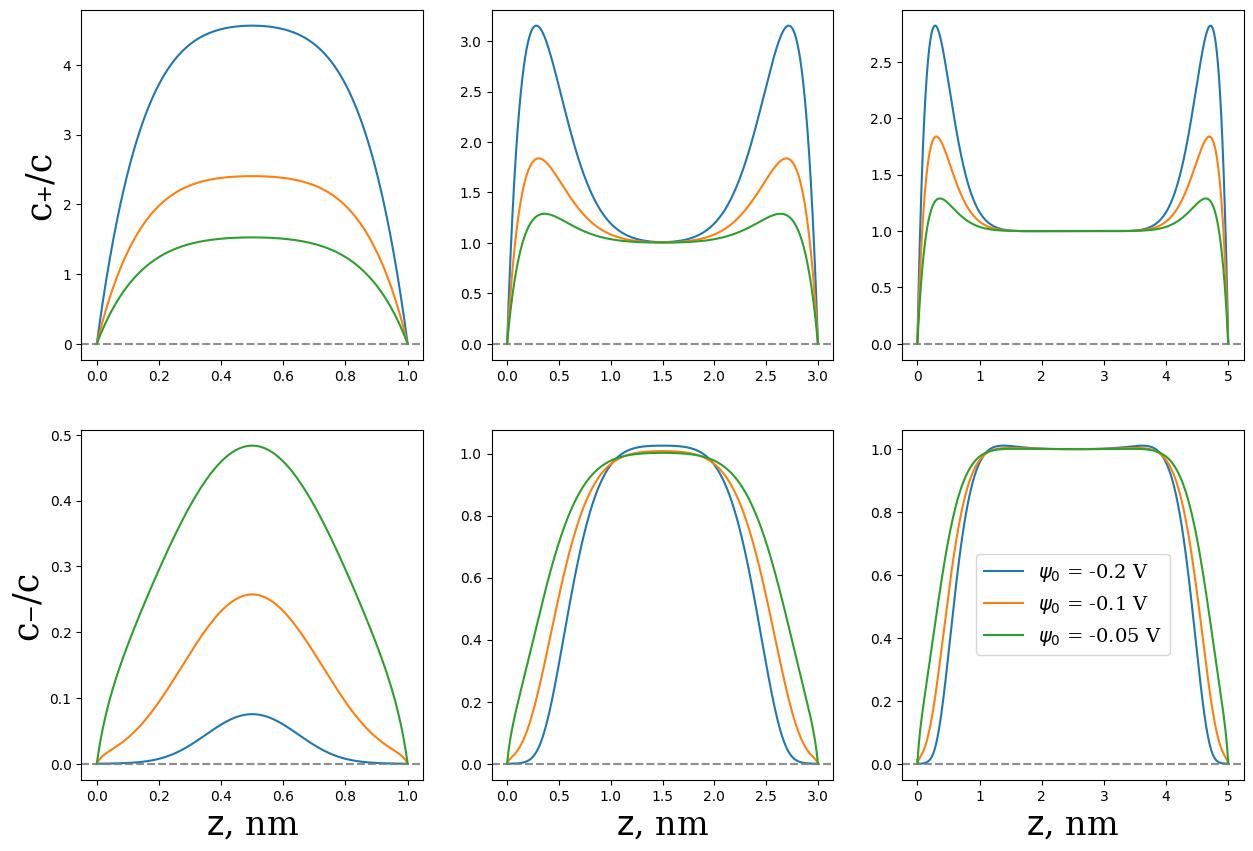}
  \caption{Dimensionless ionic concentrations, $c_{\pm}/c$, corresponding to different pore thicknesses plotted for different negative surface potentials, $\psi_{0}$. Data are shown for $\chi_{++}=1$, $\chi_{--}=0.5$, $\chi_{+-}=\chi_{-+}=0.1$, $\varepsilon_{\pm} =0$.}
  \label{fig1}
\end{figure}

\subsection{Ionic concentrations}
Fig. 1 illustrates how change in widths and surface potential affect the concentration profiles of cations and anions. At narrow pore widths, both ions concentrate predominantly at the middle of the pore, resulting in the absence of an electric double layer (EDL) at the ES/wall boundary. However, as the surface potential increases, the cation concentration increases and the anion concentration decreases due to electrostatic attraction and repulsion, respectively. When the pore width increases, electric double layers begin to form near the walls. At a width of approximately 5 nm, the EDLs become practically isolated, creating an electrically neutral bulk solution in the middle of the pore. It is important to note that positive surface potentials produce similar concentration behaviors.

\begin{figure}[ht]
  \centering
  \includegraphics[width=0.6\linewidth]{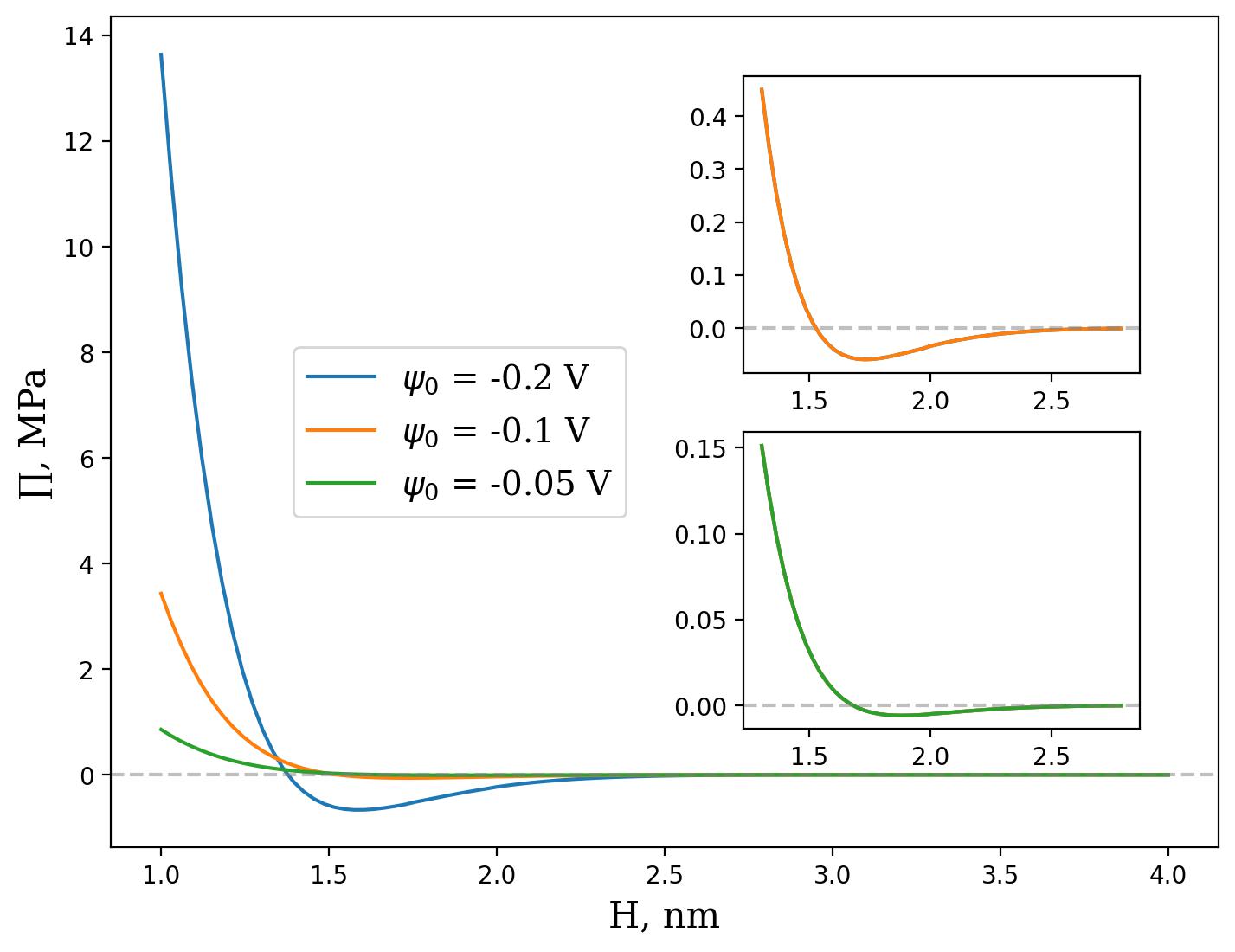}
  \caption{Disjoining pressure dependences on the pore thickness plotted for different surface potentials, $\psi_0$. Data are shown for $\chi_{++}=1$, $\chi_{--}=0.5$, $\chi_{+-}=\chi_{-+}=0.1$, $\varepsilon_{\pm} =0$.}
  \label{fig2}
\end{figure}

\begin{figure}[ht]
  \centering
  \includegraphics[width=0.8\linewidth]{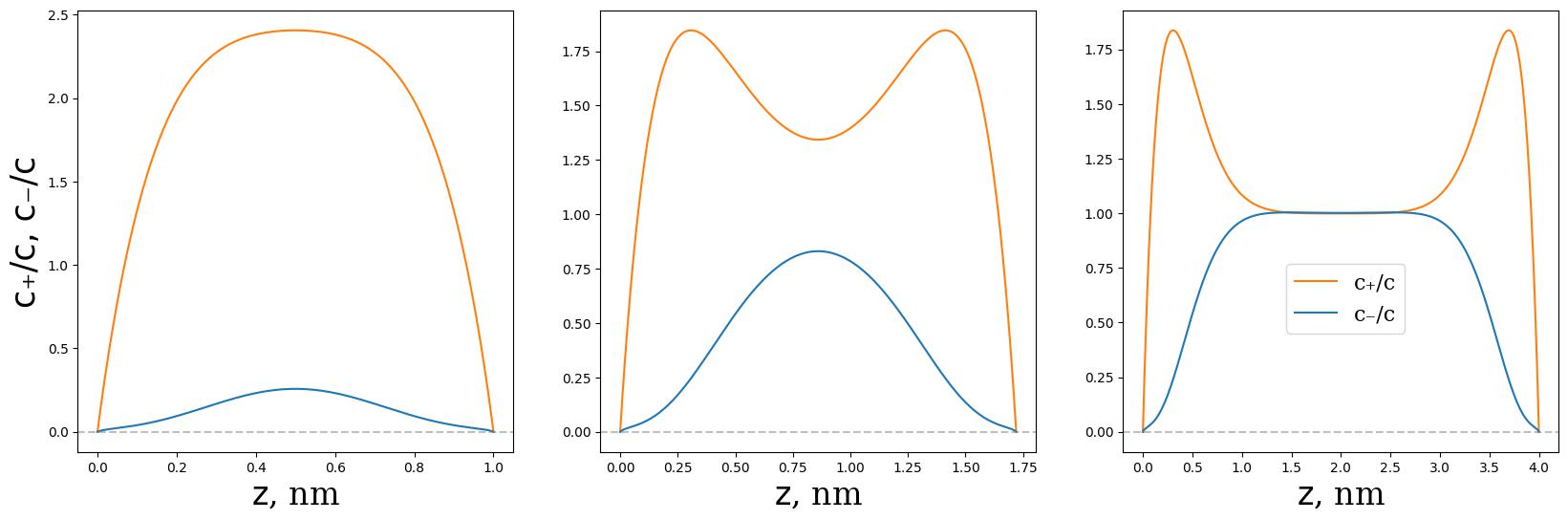}
  \caption{Ion concentration profiles at pore thicknesses up to ($H=1$~nm), at ($H=1.75$~nm), and after ($H=4$~nm) minimum disjoining pressure. Data are shown for $\chi_{++}=1$, $\chi_{--}=0.5$, $\chi_{+-}=\chi_{-+}=0.1$, $\varepsilon_{\pm} =0$, $\psi_{0}=-0.1$~V.}
  \label{fig3}
\end{figure}

\begin{figure}[ht]
  \centering
  \includegraphics[width=0.6\linewidth]{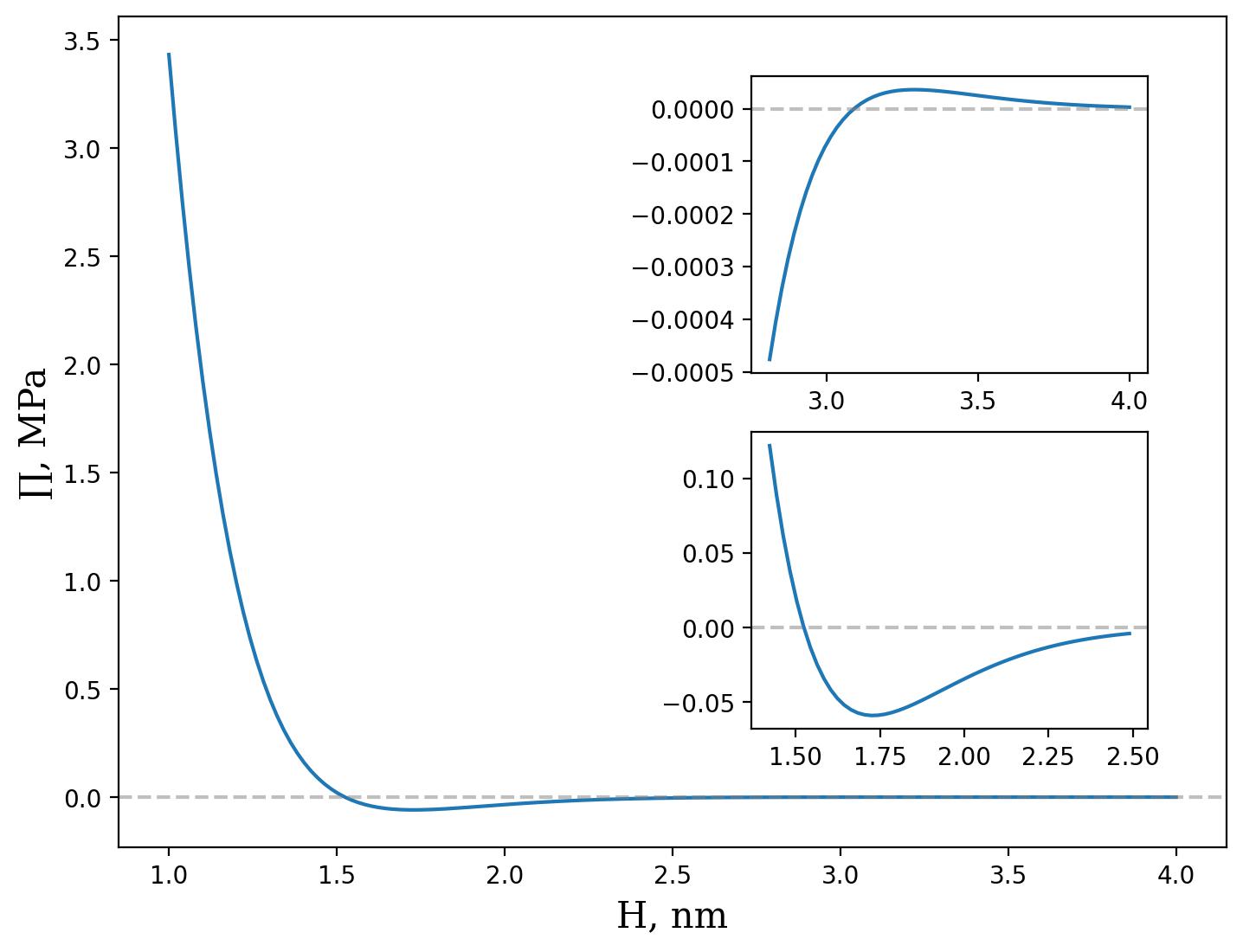}
  \caption{Disjoining pressure dependence on the pore thickness. Data are shown for $\chi_{++}=1$, $\chi_{--}=0.5$, $\chi_{+-}=\chi_{-+}=0.1$, $\varepsilon_{\pm} =0$, $\psi_{0}=-0.1$~V.}
  \label{fig4}
\end{figure}

\begin{figure}[ht]
  \centering
  \includegraphics[width=0.6\linewidth]{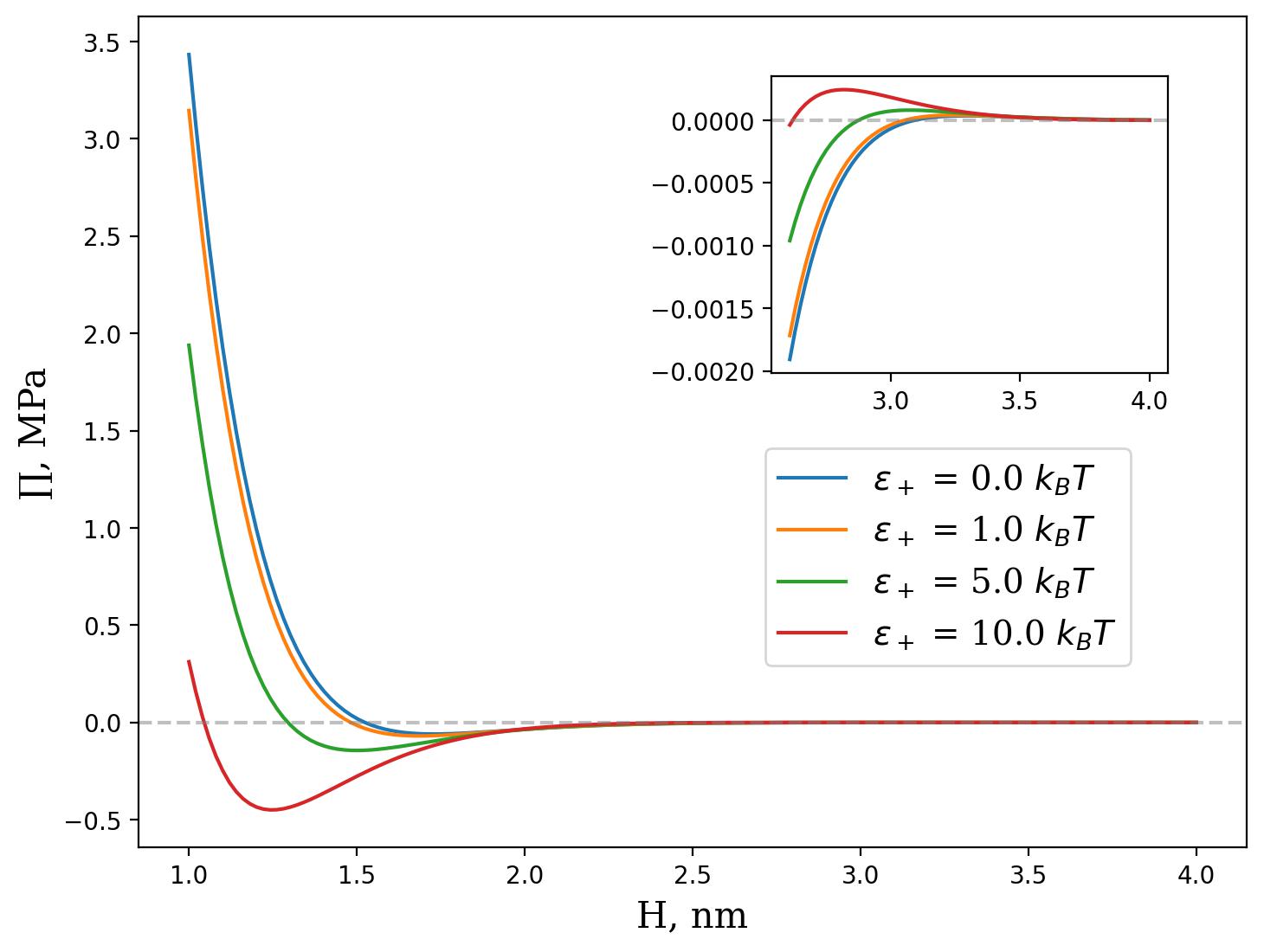}
  \caption{Disjoining pressure dependences on the pore thickness plotted for different specific adsorption parameter, $\varepsilon_+$. Data are show for $\chi_{++}=1$, $\chi_{--}=0.5$, $\chi_{+-}=\chi_{-+}=0.1$, $\varepsilon_{-}=0$, $\sigma_{+}/v^{1/3}=1$, $\psi_{0}=-0.1$~V.}
  \label{fig5}
\end{figure}

\subsection{Disjoining pressure}
In this subsection, we will examine the behavior of disjoining pressure. Fig. \ref{fig2} displays the disjoining pressure as a function of pore width at various negative surface potentials. At sufficiently small pore thickness, the disjoining pressure exhibits non-monotonic behavior, indicating effective attraction between the walls ($\Pi <0$). This minimum is more pronounced with an increase in surface potential absolute value. To comprehend the physical nature of this behavior, we will analyse the ionic concentration profiles at pore widths before ($H=1$~nm), at ($H=1.75$~nm), and after ($H=4$~nm) the minimum (see Fig. \ref{fig3}). At a narrow pore width of 1 nm, unimodal concentration profiles are present, with ions concentrated at the pore centre (see Fig. \ref{fig4}). At $H=1.75$~nm, we observe maxima in cation concentration profiles near the walls and minimum at the center of the pore. This suggests that a greater number of cations migrate from the pore center to the walls, in an attempt to form EDLs. In this configuration, there are three layers: two layers of cations near the pore walls and a layer of mixed cations and anions in between. Formation of such "sandwich" structure manifests itself in the effective "structural" attraction between the walls. At pore width of $4$~nm, as previously mentioned, we can observe isolated EDLs near the walls, with bulk solution present in the middle of the pore. In this region, the disjoining pressure exponentially damps (as shown in the upper inset in Fig. \ref{fig4}) in accordance with classical DLVO theory~\cite{derjaguin1987derjaguin,israelachvili2011intermolecular}.

We would like to state that the current model is inadequate for capturing the layering effect that occur in both room temperature ionic liquids and highly concentrated electrolyte solutions on the electrified interfaces~\cite{gurina2022disjoining,de2020interfacial,budkov2021electric} which results in the strong fluctuations on the disjoining pressure. To accurately describe these phenomena, it is necessary to take into account higher order derivative terms in the GTP~\cite{ciach2018simple} or apply nonlocal theories~\cite{de2020interfacial,brandyshev2023noether}. Nevertheless, present simple statistical theory could be used for modeling of the ES of moderate concentrations, where the ion layering is irrelevant.

\begin{figure}[ht]
  \centering
  \includegraphics[width=0.75\linewidth]{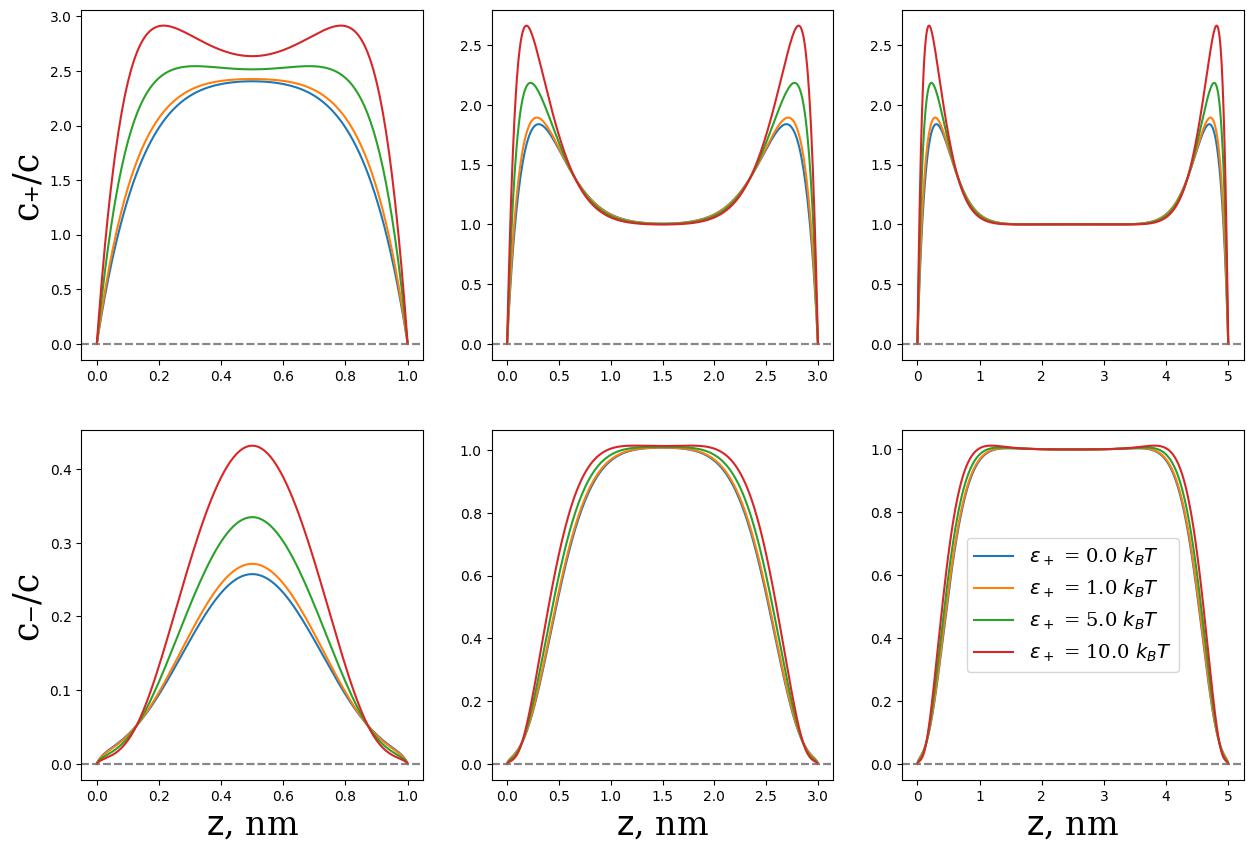}
  \caption{Ion concentration profiles corresponding to different pore thicknesses plotted for different specific adsorption parameter, $\varepsilon_+$. Data are shown for $\chi_{++}=1$, $\chi_{--}=0.5$, $\chi_{+-}=\chi_{-+}=0.1$, $\varepsilon_{-}=0$, $\sigma_{+}/v^{1/3}=1$, $\psi_{0}=-0.1$~V.}
  \label{fig6}
\end{figure}

\begin{figure}[ht]
  \centering
  \includegraphics[width=0.6\linewidth]{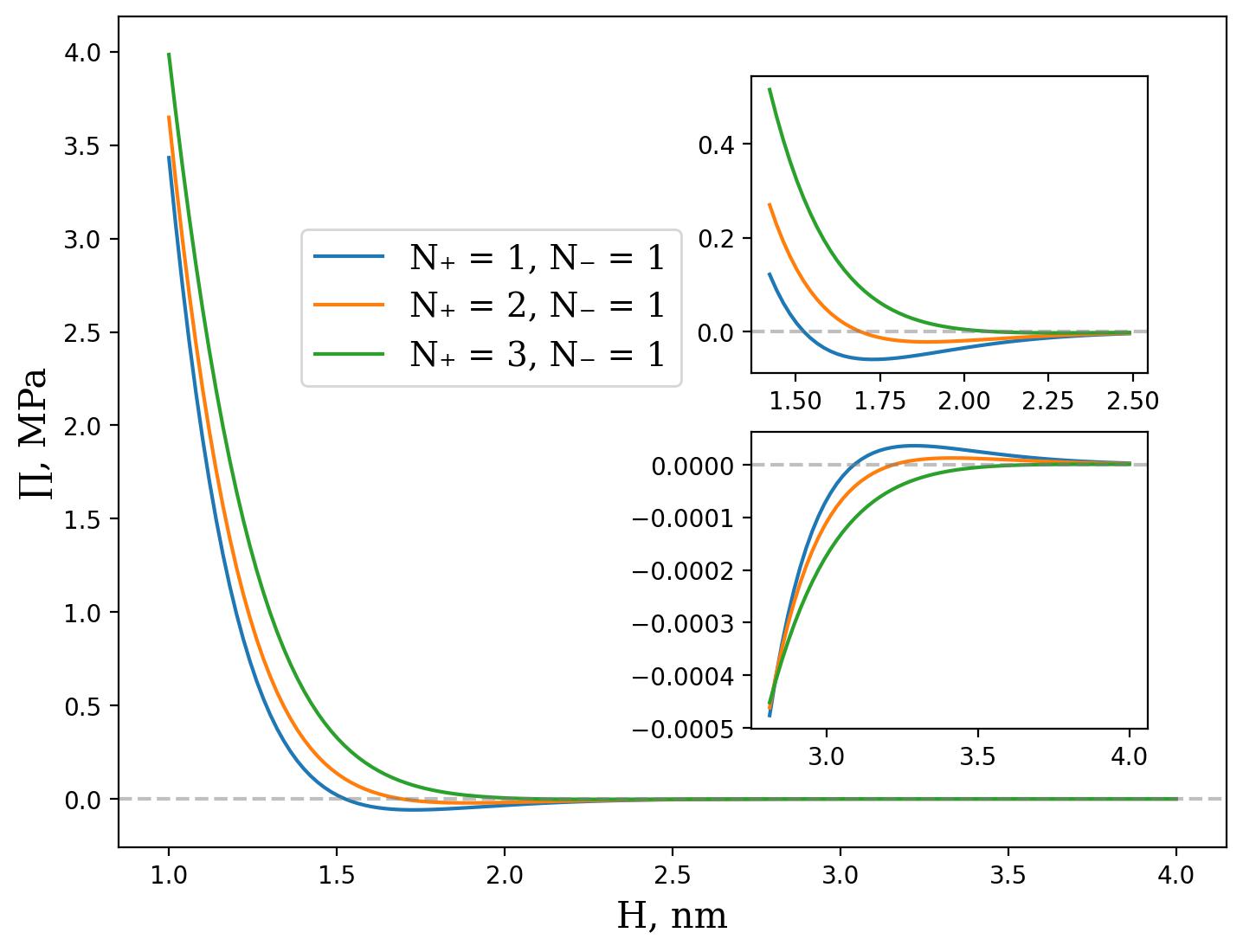}
  \caption{Disjoining pressure dependences on the pore thickness plotted for different ionic size asymmetry $N_{+} = [1, 2, 3]$ and $N_{-} = 1$. Data are shown for $\chi_{++}=1$, $\chi_{--}=0.5$, $\chi_{+-}=\chi_{-+}=0.1$, $\varepsilon_{\pm} =0$, $\sigma_{+}/v^{1/3}=1$, $\psi_{0}=-0.1$~V.}
  \label{fig7}
\end{figure}

\begin{figure}[ht]
\centering
\includegraphics[width=0.8\linewidth]{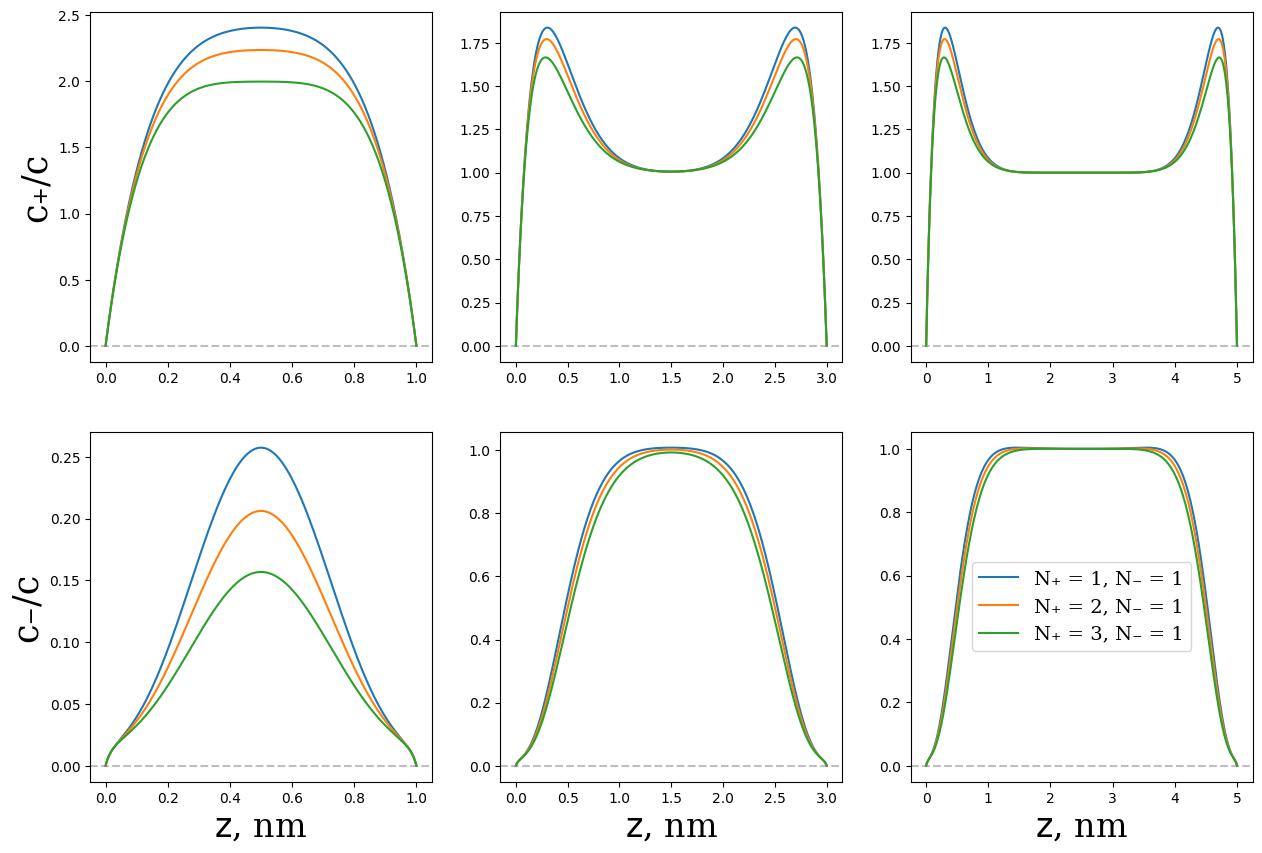}
\caption{Ion concentration profiles corresponding to different pore thicknesses plotted for different ionic size asymmetry $N_{+} = [1, 2, 3]$  and $N_{-} = 1$. Data are shown for $\chi_{++}=1$, $\chi_{--}=0.5$, $\chi_{+-}=\chi_{-+}=0.1$, $\varepsilon_{\pm} =0$, $\sigma_{+}/v^{1/3}=1$, $\psi_{0}=-0.1$~V.}
\label{fig8}
\end{figure}

\begin{figure}[ht]
  \centering
  \includegraphics[width=0.6\linewidth]{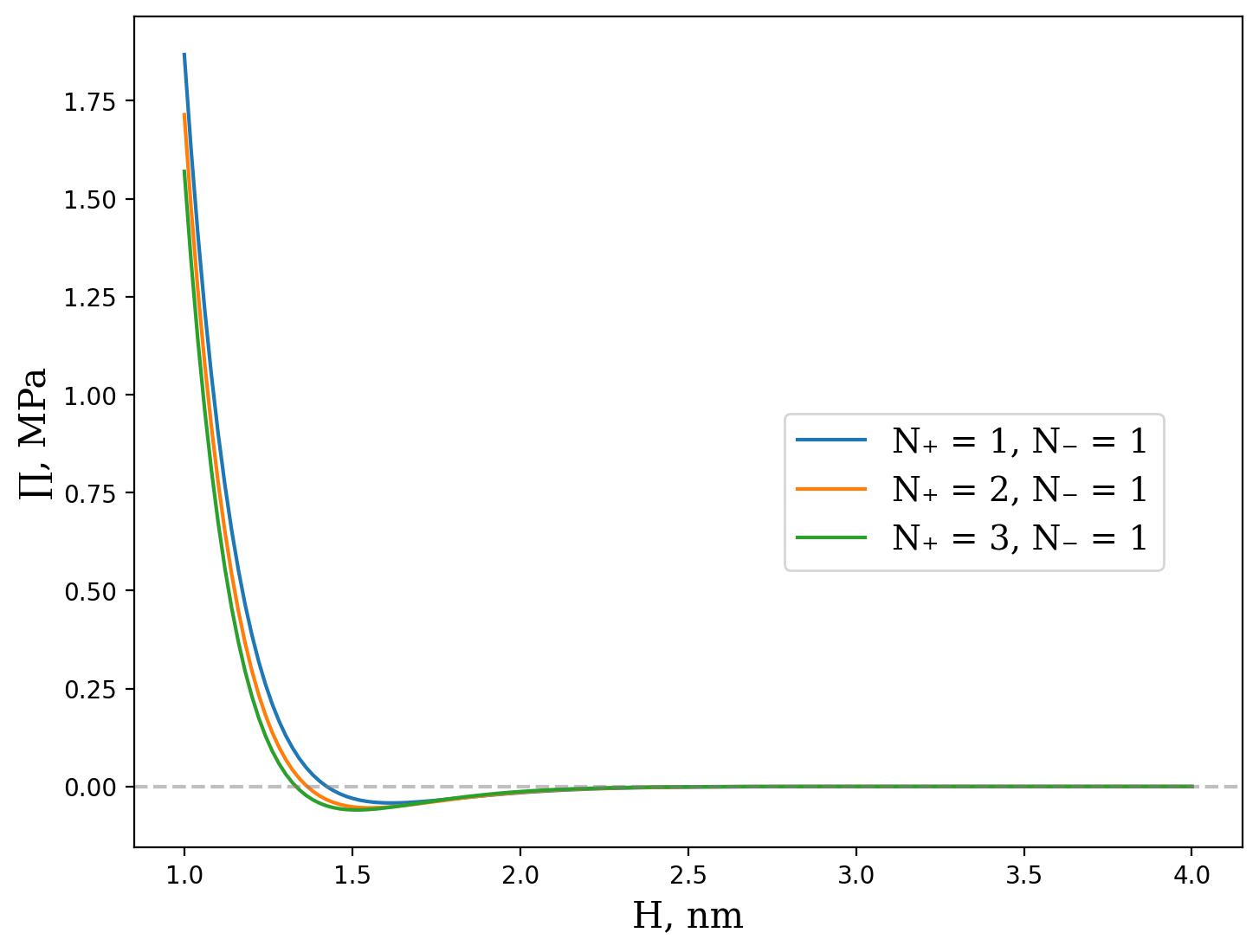}
  \caption{Disjoining pressure dependences on the pore thickness plotted for different ionic size asymmetry $N_{+} = [1, 2, 3]$ and $N_{-} = 1$. Data are shown for $\chi_{++}=1$, $\chi_{--}=0.5$, $\chi_{+-}=\chi_{-+}=0.1$, $\varepsilon_{\pm} =0$, $\sigma_{+}/v^{1/3}=1$, $\psi_{0}=0.1$~V.}
  \label{fig9}
\end{figure}

\begin{figure}[ht]
\centering
\includegraphics[width=0.8\linewidth]{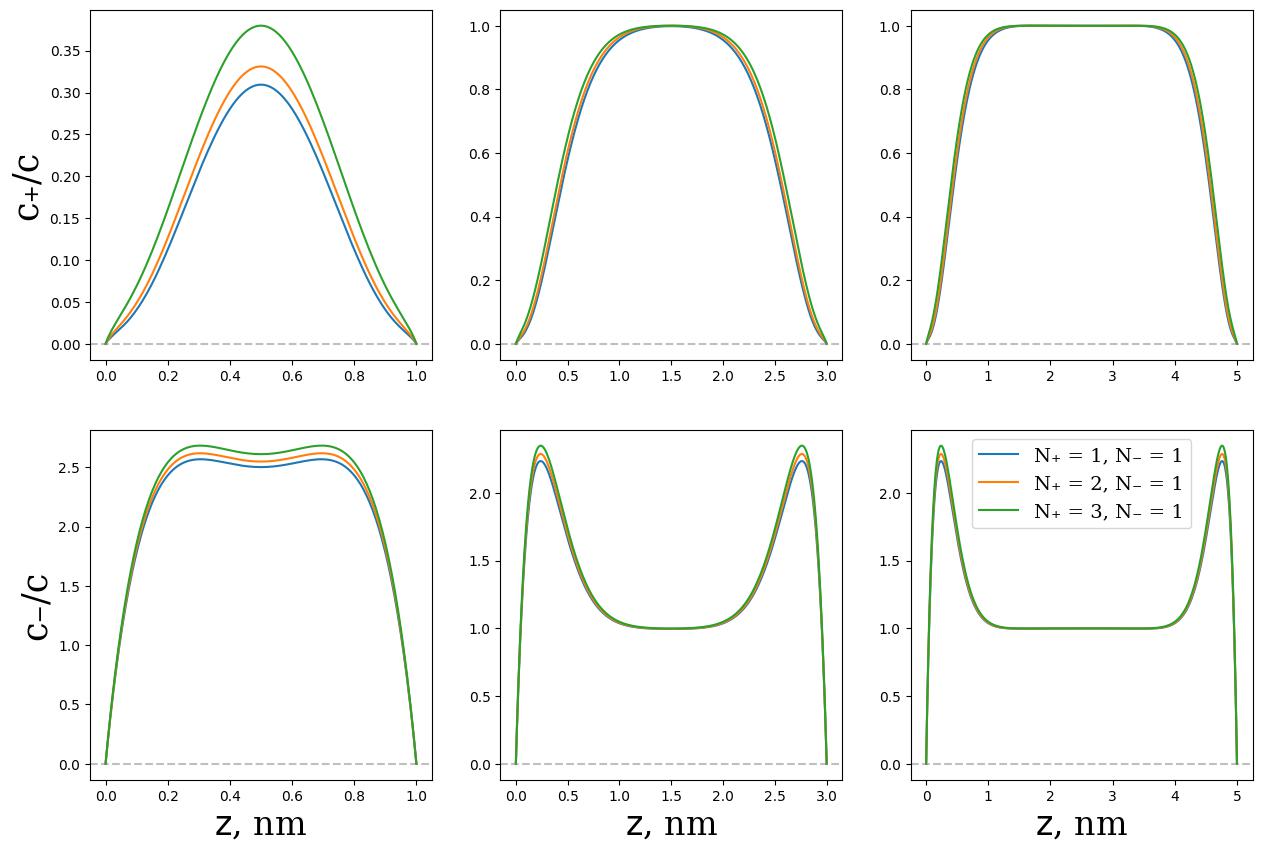}
\caption{Ion concentration profiles corresponding to different pore thicknesses plotted for different ionic size asymmetry $N_{+} = [1, 2, 3]$  and $N_{-} = 1$. Data are shown for $\chi_{++}=1$, $\chi_{--}=0.5$, $\chi_{+-}=\chi_{-+}=0.1$, $\varepsilon_{\pm} =0$, $\sigma_{+}/v^{1/3}=1$, $\psi_{0}=0.1$~V.}
\label{fig10}
\end{figure}

\subsection{Effects of specific adsorption}
It is informative to discuss how the specific adsorption of ions, described by the Gaussian-well potentials (\ref{gaussian_well}), affects the local ionic concentrations and disjoining pressure. Fig. \ref{fig5} illustrates various ionic concentration profiles for different pore thicknesses while altering the specific cation adsorption energy parameter, $\varepsilon_+$. The results demonstrate that for narrow pores ($H=1$~nm), an increase in $\varepsilon_{+}$ results in a predictable increase in the cation local concentration. However, it also leads to a substantial increase in anion local concentration due to cation-anion Coulomb attraction. With larger $H$, an increase in $\varepsilon_+$ results in sharper cation concentration profile maxima, leading to the formation of a dense part of the EDLs (Stern layers). Furthermore, $\varepsilon_{+}$ enhancement intensifies the minimum on the disjoining pressure profiles and shifts it to narrower pores. This behavior can be attributed to the fact that cation specific adsorption promotes EDL formation at smaller $H$ than is possible in the absence of specific adsorption. This is reflected in pronounced cation concentration profile maxima in Fig. \ref{fig6} at $H=1$~nm and $\varepsilon_{+}=10 k_{B}T$.

\subsection{Effects of ionic size asymmetry}
Previously, we only discussed the case of equal ionic sizes, where $N_{+}=N_{-}=1$. Now, we will briefly discuss the effects of ionic size asymmetry on disjoining pressure behavior. Fig. \ref{fig7} shows typical disjoining pressure profiles for different cation sizes, specifically $N_{+}=1$, $N_{+}=2$, and $N_{+}=3$, with a fixed anion size of $N_{-}=1$ for negatively charged pore walls ($\psi_{0}=-0.1$~V). As is seen in the figure, an increase in cation size relative to anion size practically eliminates the minimum on the disjoining pressure curve. This is because a larger cation size increases steric interactions and decreases the local cation concentration within the pore volume (see Fig. \ref{fig8}), that in turn "compensates" for the structural attraction of the walls. As in present case, according to observations made in ref.~\cite{zelko2010effects}, the electrostatic correlation of like-charge attraction between surfaces can be neutralized by increasing the diameter of counterions.

When considering walls with a positive charge ($\psi_0=0.1$ V), the size of cations has an opposite effect on the disjoining pressure - resulting in a more pronounced minimum (see Fig. \ref{fig9}). This may seem counterintuitive, but it can be attributed to the well-known phenomenon of Asakura-Oosawa depletion forces~\cite{barrat2003basic}. As cations are placed in close proximity to positively charged walls, a wall-cation depletion occurs~\cite{taboada2006electrostatic,kewalramani2016electrolyte}. The latter effect results in the expulsion of both cations and anions from the immediate vicinity of the wall surface, leading to their accumulation behind the cations. Therefore, an increase in cation effective size makes the depletion cation-wall attraction stronger. This is evident from the concentration profiles depicted in Fig. \ref{fig10} - the larger the cation size, the higher the local cation and anion concentration. We would like to mention that conventional modified Poisson-Boltzmann equations \cite{budkov2022modified,iglivc2010excluded,borukhov1997steric,kornyshev2007double,maggs2016general} do not account for depletion effects. However, incorporating structural interactions in the GTP enables us to qualitatively describe the depletion effect. It should be noted that the expulsion of counterions from the vicinity of a charged surface can be predicted using mean-field modified Poisson-Boltzmann models, provided that rotational ordering of finite water dipoles is taken into account as well~\cite{gongadze2014ions,iglivc2010excluded}. Specifically, an article~\cite{gongadze2013excluded} has demonstrated that for sufficiently high surface charge densities, the traditional trend of increasing counterion number density towards the charged surface may be reversed, resulting in a decrease in the number of counterions near the surface. This phenomenon is caused by the reduction of partial depletion of water molecules at high surface charge densities, due to the competition between the counterion Boltzmann factor and the rotationally averaged water Boltzmann factor in the closely-bound region~\cite{gongadze2013excluded}. The simulation results presented in ref.\cite{marcovitz2015water} are consistent with this conclusion. Therefore, the theoretical model must consider both the orientational ordering of finite-sized water molecules and the ionic size asymmetry, as demonstrated in~\cite{gongadze2015asymmetric}.

\section{Concluding remarks}
To summarize, we employed Cahn-Hilliard-like model of inhomogeneous electrolyte solutions that accounted for the structural and steric interactions between ions and incorporated external potentials to examine their specific interactions with pore walls. Our main focus was to observe changes in ionic concentration profiles and disjoining pressure as the pore width varied. We derived a general expression for disjoining pressure starting from the local mechanical equilibrium conditions. Our results showed that considering the structural interactions of ions resulted in a pronounced minimum on disjoining pressure profiles at small pore widths. We attributed this minimum to the formation of electric double layers on the electrified pore surfaces. Moreover, including the attractive interactions between ions and pore walls enhanced and shifted the minimum to smaller pore thicknesses. We believe that our theoretical findings could be of interest to electrochemical engineers working on supercapacitors that utilize porous electrodes impregnated with moderately concentrated electrolyte solutions.

\textbf{Data availability statement.} {\sl The data that supports the findings of this study are available within the article.}

{\bf Acknowledgements.}  
This research is supported by the Russian Science Foundation (No. 21-11-00031). The numerical calculations were partially performed on the supercomputer facilities provided by NRU HSE.

\bibliographystyle{aipnum4-2}
\bibliography{references}

%aipnum4-2.bst 2019-01-14 (MD) hand-edited version of apsrev4-1.bst
\begin{thebibliography}{57}%
\makeatletter
\providecommand \@ifxundefined [1]{%
 \@ifx{#1\undefined}
}%
\providecommand \@ifnum [1]{%
 \ifnum #1\expandafter \@firstoftwo
 \else \expandafter \@secondoftwo
 \fi
}%
\providecommand \@ifx [1]{%
 \ifx #1\expandafter \@firstoftwo
 \else \expandafter \@secondoftwo
 \fi
}%
\providecommand \natexlab [1]{#1}%
\providecommand \enquote  [1]{``#1''}%
\providecommand \bibnamefont  [1]{#1}%
\providecommand \bibfnamefont [1]{#1}%
\providecommand \citenamefont [1]{#1}%
\providecommand \href@noop [0]{\@secondoftwo}%
\providecommand \href [0]{\begingroup \@sanitize@url \@href}%
\providecommand \@href[1]{\@@startlink{#1}\@@href}%
\providecommand \@@href[1]{\endgroup#1\@@endlink}%
\providecommand \@sanitize@url [0]{\catcode `\\12\catcode `\$12\catcode
  `\&12\catcode `\#12\catcode `\^12\catcode `\_12\catcode `\%12\relax}%
\providecommand \@@startlink[1]{}%
\providecommand \@@endlink[0]{}%
\providecommand \url  [0]{\begingroup\@sanitize@url \@url }%
\providecommand \@url [1]{\endgroup\@href {#1}{\urlprefix }}%
\providecommand \urlprefix  [0]{URL }%
\providecommand \Eprint [0]{\href }%
\providecommand \doibase [0]{https://doi.org/}%
\providecommand \selectlanguage [0]{\@gobble}%
\providecommand \bibinfo  [0]{\@secondoftwo}%
\providecommand \bibfield  [0]{\@secondoftwo}%
\providecommand \translation [1]{[#1]}%
\providecommand \BibitemOpen [0]{}%
\providecommand \bibitemStop [0]{}%
\providecommand \bibitemNoStop [0]{.\EOS\space}%
\providecommand \EOS [0]{\spacefactor3000\relax}%
\providecommand \BibitemShut  [1]{\csname bibitem#1\endcsname}%
\let\auto@bib@innerbib\@empty
%</preamble>
\bibitem [{\citenamefont {Hassan}\ \emph {et~al.}(2023)\citenamefont {Hassan},
  \citenamefont {Iqbal}, \citenamefont {Gouadria}, \citenamefont {Afzal},\ and\
  \citenamefont {Hegazy}}]{hassan2023effect}%
  \BibitemOpen
  \bibfield  {author} {\bibinfo {author} {\bibfnamefont {H.}~\bibnamefont
  {Hassan}}, \bibinfo {author} {\bibfnamefont {M.~W.}\ \bibnamefont {Iqbal}},
  \bibinfo {author} {\bibfnamefont {S.}~\bibnamefont {Gouadria}}, \bibinfo
  {author} {\bibfnamefont {A.~M.}\ \bibnamefont {Afzal}},\ and\ \bibinfo
  {author} {\bibfnamefont {H.}~\bibnamefont {Hegazy}},\ }\href@noop {}
  {\bibfield  {journal} {\bibinfo  {journal} {Journal of Energy Storage}\
  }\textbf {\bibinfo {volume} {66}},\ \bibinfo {pages} {107448} (\bibinfo
  {year} {2023})}\BibitemShut {NoStop}%
\bibitem [{\citenamefont {Xia}\ \emph {et~al.}(2017)\citenamefont {Xia},
  \citenamefont {Yu}, \citenamefont {Hu},\ and\ \citenamefont
  {Chen}}]{xia2017electrolytes}%
  \BibitemOpen
  \bibfield  {author} {\bibinfo {author} {\bibfnamefont {L.}~\bibnamefont
  {Xia}}, \bibinfo {author} {\bibfnamefont {L.}~\bibnamefont {Yu}}, \bibinfo
  {author} {\bibfnamefont {D.}~\bibnamefont {Hu}},\ and\ \bibinfo {author}
  {\bibfnamefont {G.~Z.}\ \bibnamefont {Chen}},\ }\href@noop {} {\bibfield
  {journal} {\bibinfo  {journal} {Materials Chemistry Frontiers}\ }\textbf
  {\bibinfo {volume} {1}},\ \bibinfo {pages} {584} (\bibinfo {year}
  {2017})}\BibitemShut {NoStop}%
\bibitem [{\citenamefont {Zheng}\ \emph {et~al.}(2017)\citenamefont {Zheng},
  \citenamefont {Lochala}, \citenamefont {Kwok}, \citenamefont {Deng},\ and\
  \citenamefont {Xiao}}]{zheng2017research}%
  \BibitemOpen
  \bibfield  {author} {\bibinfo {author} {\bibfnamefont {J.}~\bibnamefont
  {Zheng}}, \bibinfo {author} {\bibfnamefont {J.~A.}\ \bibnamefont {Lochala}},
  \bibinfo {author} {\bibfnamefont {A.}~\bibnamefont {Kwok}}, \bibinfo {author}
  {\bibfnamefont {Z.~D.}\ \bibnamefont {Deng}},\ and\ \bibinfo {author}
  {\bibfnamefont {J.}~\bibnamefont {Xiao}},\ }\href@noop {} {\bibfield
  {journal} {\bibinfo  {journal} {Advanced Science}\ }\textbf {\bibinfo
  {volume} {4}},\ \bibinfo {pages} {1700032} (\bibinfo {year}
  {2017})}\BibitemShut {NoStop}%
\bibitem [{\citenamefont {Xu}\ \emph {et~al.}(2023)\citenamefont {Xu},
  \citenamefont {Mao}, \citenamefont {Qu}, \citenamefont {Wang}, \citenamefont
  {Yu}, \citenamefont {Luo}, \citenamefont {Shi}, \citenamefont {Mao},
  \citenamefont {Ding},\ and\ \citenamefont {Liu}}]{xu2023electrochemical}%
  \BibitemOpen
  \bibfield  {author} {\bibinfo {author} {\bibfnamefont {Y.}~\bibnamefont
  {Xu}}, \bibinfo {author} {\bibfnamefont {Z.}~\bibnamefont {Mao}}, \bibinfo
  {author} {\bibfnamefont {R.}~\bibnamefont {Qu}}, \bibinfo {author}
  {\bibfnamefont {J.}~\bibnamefont {Wang}}, \bibinfo {author} {\bibfnamefont
  {J.}~\bibnamefont {Yu}}, \bibinfo {author} {\bibfnamefont {X.}~\bibnamefont
  {Luo}}, \bibinfo {author} {\bibfnamefont {M.}~\bibnamefont {Shi}}, \bibinfo
  {author} {\bibfnamefont {X.}~\bibnamefont {Mao}}, \bibinfo {author}
  {\bibfnamefont {J.}~\bibnamefont {Ding}},\ and\ \bibinfo {author}
  {\bibfnamefont {B.}~\bibnamefont {Liu}},\ }\href@noop {} {\bibfield
  {journal} {\bibinfo  {journal} {Nature Water}\ ,\ \bibinfo {pages} {1}}
  (\bibinfo {year} {2023})}\BibitemShut {NoStop}%
\bibitem [{\citenamefont {Sharma}, \citenamefont {Borkute},\ and\ \citenamefont
  {Gumfekar}(2022)}]{sharma2022biomimetic}%
  \BibitemOpen
  \bibfield  {author} {\bibinfo {author} {\bibfnamefont {V.}~\bibnamefont
  {Sharma}}, \bibinfo {author} {\bibfnamefont {G.}~\bibnamefont {Borkute}},\
  and\ \bibinfo {author} {\bibfnamefont {S.~P.}\ \bibnamefont {Gumfekar}},\
  }\href@noop {} {\bibfield  {journal} {\bibinfo  {journal} {Chemical
  Engineering Journal}\ }\textbf {\bibinfo {volume} {433}},\ \bibinfo {pages}
  {133823} (\bibinfo {year} {2022})}\BibitemShut {NoStop}%
\bibitem [{\citenamefont {Aluru}\ \emph {et~al.}(2023)\citenamefont {Aluru},
  \citenamefont {Aydin}, \citenamefont {Bazant}, \citenamefont {Blankschtein},
  \citenamefont {Brozena}, \citenamefont {de~Souza}, \citenamefont {Elimelech},
  \citenamefont {Faucher}, \citenamefont {Fourkas}, \citenamefont {Koman} \emph
  {et~al.}}]{aluru2023fluids}%
  \BibitemOpen
  \bibfield  {author} {\bibinfo {author} {\bibfnamefont {N.~R.}\ \bibnamefont
  {Aluru}}, \bibinfo {author} {\bibfnamefont {F.}~\bibnamefont {Aydin}},
  \bibinfo {author} {\bibfnamefont {M.~Z.}\ \bibnamefont {Bazant}}, \bibinfo
  {author} {\bibfnamefont {D.}~\bibnamefont {Blankschtein}}, \bibinfo {author}
  {\bibfnamefont {A.~H.}\ \bibnamefont {Brozena}}, \bibinfo {author}
  {\bibfnamefont {J.~P.}\ \bibnamefont {de~Souza}}, \bibinfo {author}
  {\bibfnamefont {M.}~\bibnamefont {Elimelech}}, \bibinfo {author}
  {\bibfnamefont {S.}~\bibnamefont {Faucher}}, \bibinfo {author} {\bibfnamefont
  {J.~T.}\ \bibnamefont {Fourkas}}, \bibinfo {author} {\bibfnamefont {V.~B.}\
  \bibnamefont {Koman}}, \emph {et~al.},\ }\href@noop {} {\bibfield  {journal}
  {\bibinfo  {journal} {Chemical reviews}\ }\textbf {\bibinfo {volume} {123}},\
  \bibinfo {pages} {2737} (\bibinfo {year} {2023})}\BibitemShut {NoStop}%
\bibitem [{\citenamefont {Shin}, \citenamefont {Guiver},\ and\ \citenamefont
  {Lee}(2017)}]{shin2017hydrocarbon}%
  \BibitemOpen
  \bibfield  {author} {\bibinfo {author} {\bibfnamefont {D.~W.}\ \bibnamefont
  {Shin}}, \bibinfo {author} {\bibfnamefont {M.~D.}\ \bibnamefont {Guiver}},\
  and\ \bibinfo {author} {\bibfnamefont {Y.~M.}\ \bibnamefont {Lee}},\
  }\href@noop {} {\bibfield  {journal} {\bibinfo  {journal} {Chemical reviews}\
  }\textbf {\bibinfo {volume} {117}},\ \bibinfo {pages} {4759} (\bibinfo {year}
  {2017})}\BibitemShut {NoStop}%
\bibitem [{\citenamefont {Luo}, \citenamefont {Abdu},\ and\ \citenamefont
  {Wessling}(2018)}]{luo2018selectivity}%
  \BibitemOpen
  \bibfield  {author} {\bibinfo {author} {\bibfnamefont {T.}~\bibnamefont
  {Luo}}, \bibinfo {author} {\bibfnamefont {S.}~\bibnamefont {Abdu}},\ and\
  \bibinfo {author} {\bibfnamefont {M.}~\bibnamefont {Wessling}},\ }\href@noop
  {} {\bibfield  {journal} {\bibinfo  {journal} {Journal of membrane science}\
  }\textbf {\bibinfo {volume} {555}},\ \bibinfo {pages} {429} (\bibinfo {year}
  {2018})}\BibitemShut {NoStop}%
\bibitem [{\citenamefont {Ben-Yaakov}, \citenamefont {Andelman},\ and\
  \citenamefont {Podgornik}(2011)}]{ben2011dielectric}%
  \BibitemOpen
  \bibfield  {author} {\bibinfo {author} {\bibfnamefont {D.}~\bibnamefont
  {Ben-Yaakov}}, \bibinfo {author} {\bibfnamefont {D.}~\bibnamefont
  {Andelman}},\ and\ \bibinfo {author} {\bibfnamefont {R.}~\bibnamefont
  {Podgornik}},\ }\href@noop {} {\bibfield  {journal} {\bibinfo  {journal} {The
  Journal of chemical physics}\ }\textbf {\bibinfo {volume} {134}},\ \bibinfo
  {pages} {074705} (\bibinfo {year} {2011})}\BibitemShut {NoStop}%
\bibitem [{\citenamefont {Frydel}(2011)}]{frydel2011polarizable}%
  \BibitemOpen
  \bibfield  {author} {\bibinfo {author} {\bibfnamefont {D.}~\bibnamefont
  {Frydel}},\ }\href@noop {} {\bibfield  {journal} {\bibinfo  {journal} {The
  Journal of chemical physics}\ }\textbf {\bibinfo {volume} {134}},\ \bibinfo
  {pages} {234704} (\bibinfo {year} {2011})}\BibitemShut {NoStop}%
\bibitem [{\citenamefont {Hatlo}, \citenamefont {Van~Roij},\ and\ \citenamefont
  {Lue}(2012)}]{hatlo2012electric}%
  \BibitemOpen
  \bibfield  {author} {\bibinfo {author} {\bibfnamefont {M.~M.}\ \bibnamefont
  {Hatlo}}, \bibinfo {author} {\bibfnamefont {R.}~\bibnamefont {Van~Roij}},\
  and\ \bibinfo {author} {\bibfnamefont {L.}~\bibnamefont {Lue}},\ }\href@noop
  {} {\bibfield  {journal} {\bibinfo  {journal} {Europhysics Letters}\ }\textbf
  {\bibinfo {volume} {97}},\ \bibinfo {pages} {28010} (\bibinfo {year}
  {2012})}\BibitemShut {NoStop}%
\bibitem [{\citenamefont {Goodwin}, \citenamefont {Feng},\ and\ \citenamefont
  {Kornyshev}(2017)}]{goodwin2017mean}%
  \BibitemOpen
  \bibfield  {author} {\bibinfo {author} {\bibfnamefont {Z.~A.}\ \bibnamefont
  {Goodwin}}, \bibinfo {author} {\bibfnamefont {G.}~\bibnamefont {Feng}},\ and\
  \bibinfo {author} {\bibfnamefont {A.~A.}\ \bibnamefont {Kornyshev}},\
  }\href@noop {} {\bibfield  {journal} {\bibinfo  {journal} {Electrochimica
  Acta}\ }\textbf {\bibinfo {volume} {225}},\ \bibinfo {pages} {190} (\bibinfo
  {year} {2017})}\BibitemShut {NoStop}%
\bibitem [{\citenamefont {Uematsu}, \citenamefont {Netz},\ and\ \citenamefont
  {Bonthuis}(2018)}]{uematsu2018effects}%
  \BibitemOpen
  \bibfield  {author} {\bibinfo {author} {\bibfnamefont {Y.}~\bibnamefont
  {Uematsu}}, \bibinfo {author} {\bibfnamefont {R.~R.}\ \bibnamefont {Netz}},\
  and\ \bibinfo {author} {\bibfnamefont {D.~J.}\ \bibnamefont {Bonthuis}},\
  }\href@noop {} {\bibfield  {journal} {\bibinfo  {journal} {Journal of
  Physics: Condensed Matter}\ }\textbf {\bibinfo {volume} {30}},\ \bibinfo
  {pages} {064002} (\bibinfo {year} {2018})}\BibitemShut {NoStop}%
\bibitem [{\citenamefont {Budkov}, \citenamefont {Zavarzin},\ and\
  \citenamefont {Kolesnikov}(2021)}]{budkov2021theory}%
  \BibitemOpen
  \bibfield  {author} {\bibinfo {author} {\bibfnamefont {Y.~A.}\ \bibnamefont
  {Budkov}}, \bibinfo {author} {\bibfnamefont {S.~V.}\ \bibnamefont
  {Zavarzin}},\ and\ \bibinfo {author} {\bibfnamefont {A.~L.}\ \bibnamefont
  {Kolesnikov}},\ }\href@noop {} {\bibfield  {journal} {\bibinfo  {journal}
  {The Journal of Physical Chemistry C}\ }\textbf {\bibinfo {volume} {125}},\
  \bibinfo {pages} {21151} (\bibinfo {year} {2021})}\BibitemShut {NoStop}%
\bibitem [{\citenamefont {Budkov}\ \emph {et~al.}(2020)\citenamefont {Budkov},
  \citenamefont {Sergeev}, \citenamefont {Zavarzin},\ and\ \citenamefont
  {Kolesnikov}}]{budkov2020two}%
  \BibitemOpen
  \bibfield  {author} {\bibinfo {author} {\bibfnamefont {Y.~A.}\ \bibnamefont
  {Budkov}}, \bibinfo {author} {\bibfnamefont {A.~V.}\ \bibnamefont {Sergeev}},
  \bibinfo {author} {\bibfnamefont {S.~V.}\ \bibnamefont {Zavarzin}},\ and\
  \bibinfo {author} {\bibfnamefont {A.~L.}\ \bibnamefont {Kolesnikov}},\
  }\href@noop {} {\bibfield  {journal} {\bibinfo  {journal} {The Journal of
  Physical Chemistry C}\ }\textbf {\bibinfo {volume} {124}},\ \bibinfo {pages}
  {16308} (\bibinfo {year} {2020})}\BibitemShut {NoStop}%
\bibitem [{\citenamefont {Podgornik}(2018)}]{podgornik}%
  \BibitemOpen
  \bibfield  {author} {\bibinfo {author} {\bibfnamefont {R.}~\bibnamefont
  {Podgornik}},\ }\href@noop {} {\bibfield  {journal} {\bibinfo  {journal} {The
  Journal of chemical physics}\ }\textbf {\bibinfo {volume} {149}},\ \bibinfo
  {pages} {104701} (\bibinfo {year} {2018})}\BibitemShut {NoStop}%
\bibitem [{\citenamefont {Naji}\ \emph {et~al.}(2013)\citenamefont {Naji},
  \citenamefont {Kandu{\v{c}}}, \citenamefont {Forsman},\ and\ \citenamefont
  {Podgornik}}]{naji2013perspective}%
  \BibitemOpen
  \bibfield  {author} {\bibinfo {author} {\bibfnamefont {A.}~\bibnamefont
  {Naji}}, \bibinfo {author} {\bibfnamefont {M.}~\bibnamefont {Kandu{\v{c}}}},
  \bibinfo {author} {\bibfnamefont {J.}~\bibnamefont {Forsman}},\ and\ \bibinfo
  {author} {\bibfnamefont {R.}~\bibnamefont {Podgornik}},\ }\href@noop {}
  {\bibfield  {journal} {\bibinfo  {journal} {The Journal of chemical physics}\
  }\textbf {\bibinfo {volume} {139}},\ \bibinfo {pages} {150901} (\bibinfo
  {year} {2013})}\BibitemShut {NoStop}%
\bibitem [{\citenamefont {Blossey}(2023)}]{blossey2023poisson}%
  \BibitemOpen
  \bibfield  {author} {\bibinfo {author} {\bibfnamefont {R.}~\bibnamefont
  {Blossey}},\ }in\ \href@noop {} {\emph {\bibinfo {booktitle} {The
  Poisson-Boltzmann Equation: An Introduction}}}\ (\bibinfo  {publisher}
  {Springer},\ \bibinfo {year} {2023})\ pp.\ \bibinfo {pages}
  {53--96}\BibitemShut {NoStop}%
\bibitem [{\citenamefont {Budkov}\ and\ \citenamefont
  {Kolesnikov}(2021)}]{budkov2021electric}%
  \BibitemOpen
  \bibfield  {author} {\bibinfo {author} {\bibfnamefont {Y.~A.}\ \bibnamefont
  {Budkov}}\ and\ \bibinfo {author} {\bibfnamefont {A.~L.}\ \bibnamefont
  {Kolesnikov}},\ }\href@noop {} {\bibfield  {journal} {\bibinfo  {journal}
  {Current Opinion in Electrochemistry}\ }\textbf {\bibinfo {volume} {33}},\
  \bibinfo {pages} {100931} (\bibinfo {year} {2021})}\BibitemShut {NoStop}%
\bibitem [{\citenamefont {Misra}\ \emph {et~al.}(2019)\citenamefont {Misra},
  \citenamefont {de~Souza}, \citenamefont {Blankschtein},\ and\ \citenamefont
  {Bazant}}]{misra2019theory}%
  \BibitemOpen
  \bibfield  {author} {\bibinfo {author} {\bibfnamefont {R.~P.}\ \bibnamefont
  {Misra}}, \bibinfo {author} {\bibfnamefont {J.~P.}\ \bibnamefont {de~Souza}},
  \bibinfo {author} {\bibfnamefont {D.}~\bibnamefont {Blankschtein}},\ and\
  \bibinfo {author} {\bibfnamefont {M.~Z.}\ \bibnamefont {Bazant}},\
  }\href@noop {} {\bibfield  {journal} {\bibinfo  {journal} {Langmuir}\
  }\textbf {\bibinfo {volume} {35}},\ \bibinfo {pages} {11550} (\bibinfo {year}
  {2019})}\BibitemShut {NoStop}%
\bibitem [{\citenamefont {de~Souza}\ and\ \citenamefont
  {Bazant}(2020)}]{de2020continuum}%
  \BibitemOpen
  \bibfield  {author} {\bibinfo {author} {\bibfnamefont {J.~P.}\ \bibnamefont
  {de~Souza}}\ and\ \bibinfo {author} {\bibfnamefont {M.~Z.}\ \bibnamefont
  {Bazant}},\ }\href@noop {} {\bibfield  {journal} {\bibinfo  {journal} {The
  Journal of Physical Chemistry C}\ }\textbf {\bibinfo {volume} {124}},\
  \bibinfo {pages} {11414} (\bibinfo {year} {2020})}\BibitemShut {NoStop}%
\bibitem [{\citenamefont {de~Souza}\ \emph {et~al.}(2020)\citenamefont
  {de~Souza}, \citenamefont {Goodwin}, \citenamefont {McEldrew}, \citenamefont
  {Kornyshev},\ and\ \citenamefont {Bazant}}]{de2020interfacial}%
  \BibitemOpen
  \bibfield  {author} {\bibinfo {author} {\bibfnamefont {J.~P.}\ \bibnamefont
  {de~Souza}}, \bibinfo {author} {\bibfnamefont {Z.~A.}\ \bibnamefont
  {Goodwin}}, \bibinfo {author} {\bibfnamefont {M.}~\bibnamefont {McEldrew}},
  \bibinfo {author} {\bibfnamefont {A.~A.}\ \bibnamefont {Kornyshev}},\ and\
  \bibinfo {author} {\bibfnamefont {M.~Z.}\ \bibnamefont {Bazant}},\
  }\href@noop {} {\bibfield  {journal} {\bibinfo  {journal} {Physical Review
  Letters}\ }\textbf {\bibinfo {volume} {125}},\ \bibinfo {pages} {116001}
  (\bibinfo {year} {2020})}\BibitemShut {NoStop}%
\bibitem [{\citenamefont {Shi}\ \emph {et~al.}(2023)\citenamefont {Shi},
  \citenamefont {Smith}, \citenamefont {Santiso},\ and\ \citenamefont
  {Gubbins}}]{shi2023perspective}%
  \BibitemOpen
  \bibfield  {author} {\bibinfo {author} {\bibfnamefont {K.}~\bibnamefont
  {Shi}}, \bibinfo {author} {\bibfnamefont {E.~R.}\ \bibnamefont {Smith}},
  \bibinfo {author} {\bibfnamefont {E.~E.}\ \bibnamefont {Santiso}},\ and\
  \bibinfo {author} {\bibfnamefont {K.~E.}\ \bibnamefont {Gubbins}},\
  }\href@noop {} {\bibfield  {journal} {\bibinfo  {journal} {The Journal of
  Chemical Physics}\ }\textbf {\bibinfo {volume} {158}},\ \bibinfo {pages}
  {040901} (\bibinfo {year} {2023})}\BibitemShut {NoStop}%
\bibitem [{\citenamefont {Kolesnikov}, \citenamefont {Budkov},\ and\
  \citenamefont {Gor}(2021)}]{kolesnikov2021models}%
  \BibitemOpen
  \bibfield  {author} {\bibinfo {author} {\bibfnamefont {A.}~\bibnamefont
  {Kolesnikov}}, \bibinfo {author} {\bibfnamefont {Y.~A.}\ \bibnamefont
  {Budkov}},\ and\ \bibinfo {author} {\bibfnamefont {G.}~\bibnamefont {Gor}},\
  }\href@noop {} {\bibfield  {journal} {\bibinfo  {journal} {Journal of
  Physics: Condensed Matter}\ }\textbf {\bibinfo {volume} {34}},\ \bibinfo
  {pages} {063002} (\bibinfo {year} {2021})}\BibitemShut {NoStop}%
\bibitem [{\citenamefont {Gurina}\ \emph {et~al.}(2022)\citenamefont {Gurina},
  \citenamefont {Odintsova}, \citenamefont {Kolesnikov}, \citenamefont
  {Kiselev},\ and\ \citenamefont {Budkov}}]{gurina2022disjoining}%
  \BibitemOpen
  \bibfield  {author} {\bibinfo {author} {\bibfnamefont {D.}~\bibnamefont
  {Gurina}}, \bibinfo {author} {\bibfnamefont {E.}~\bibnamefont {Odintsova}},
  \bibinfo {author} {\bibfnamefont {A.}~\bibnamefont {Kolesnikov}}, \bibinfo
  {author} {\bibfnamefont {M.}~\bibnamefont {Kiselev}},\ and\ \bibinfo {author}
  {\bibfnamefont {Y.}~\bibnamefont {Budkov}},\ }\href@noop {} {\bibfield
  {journal} {\bibinfo  {journal} {Journal of Molecular Liquids}\ }\textbf
  {\bibinfo {volume} {366}},\ \bibinfo {pages} {120307} (\bibinfo {year}
  {2022})}\BibitemShut {NoStop}%
\bibitem [{\citenamefont {Kolesnikov}, \citenamefont {Mazur},\ and\
  \citenamefont {Budkov}(2022)}]{kolesnikov2022electrosorption}%
  \BibitemOpen
  \bibfield  {author} {\bibinfo {author} {\bibfnamefont {A.~L.}\ \bibnamefont
  {Kolesnikov}}, \bibinfo {author} {\bibfnamefont {D.~A.}\ \bibnamefont
  {Mazur}},\ and\ \bibinfo {author} {\bibfnamefont {Y.~A.}\ \bibnamefont
  {Budkov}},\ }\href@noop {} {\bibfield  {journal} {\bibinfo  {journal}
  {Europhysics Letters}\ }\textbf {\bibinfo {volume} {140}},\ \bibinfo {pages}
  {16001} (\bibinfo {year} {2022})}\BibitemShut {NoStop}%
\bibitem [{\citenamefont {Koczwara}\ \emph {et~al.}(2017)\citenamefont
  {Koczwara}, \citenamefont {Rumswinkel}, \citenamefont {Prehal}, \citenamefont
  {Jackel}, \citenamefont {Elsasser}, \citenamefont {Amenitsch}, \citenamefont
  {Presser}, \citenamefont {Husing},\ and\ \citenamefont
  {Paris}}]{koczwara2017situ}%
  \BibitemOpen
  \bibfield  {author} {\bibinfo {author} {\bibfnamefont {C.}~\bibnamefont
  {Koczwara}}, \bibinfo {author} {\bibfnamefont {S.}~\bibnamefont
  {Rumswinkel}}, \bibinfo {author} {\bibfnamefont {C.}~\bibnamefont {Prehal}},
  \bibinfo {author} {\bibfnamefont {N.}~\bibnamefont {Jackel}}, \bibinfo
  {author} {\bibfnamefont {M.~S.}\ \bibnamefont {Elsasser}}, \bibinfo {author}
  {\bibfnamefont {H.}~\bibnamefont {Amenitsch}}, \bibinfo {author}
  {\bibfnamefont {V.}~\bibnamefont {Presser}}, \bibinfo {author} {\bibfnamefont
  {N.}~\bibnamefont {Husing}},\ and\ \bibinfo {author} {\bibfnamefont
  {O.}~\bibnamefont {Paris}},\ }\href@noop {} {\bibfield  {journal} {\bibinfo
  {journal} {ACS applied materials \& interfaces}\ }\textbf {\bibinfo {volume}
  {9}},\ \bibinfo {pages} {23319} (\bibinfo {year} {2017})}\BibitemShut
  {NoStop}%
\bibitem [{\citenamefont {Chen}\ \emph {et~al.}(2022)\citenamefont {Chen},
  \citenamefont {Danilov}, \citenamefont {Eichel},\ and\ \citenamefont
  {Notten}}]{chen2022porous}%
  \BibitemOpen
  \bibfield  {author} {\bibinfo {author} {\bibfnamefont {Z.}~\bibnamefont
  {Chen}}, \bibinfo {author} {\bibfnamefont {D.~L.}\ \bibnamefont {Danilov}},
  \bibinfo {author} {\bibfnamefont {R.-A.}\ \bibnamefont {Eichel}},\ and\
  \bibinfo {author} {\bibfnamefont {P.~H.}\ \bibnamefont {Notten}},\
  }\href@noop {} {\bibfield  {journal} {\bibinfo  {journal} {Advanced Energy
  Materials}\ }\textbf {\bibinfo {volume} {12}},\ \bibinfo {pages} {2201506}
  (\bibinfo {year} {2022})}\BibitemShut {NoStop}%
\bibitem [{\citenamefont {Da~Silva}\ \emph {et~al.}(2020)\citenamefont
  {Da~Silva}, \citenamefont {Cesar}, \citenamefont {Moreira}, \citenamefont
  {Santos}, \citenamefont {De~Souza}, \citenamefont {Pires}, \citenamefont
  {Vicentini}, \citenamefont {Nunes},\ and\ \citenamefont
  {Zanin}}]{da2020reviewing}%
  \BibitemOpen
  \bibfield  {author} {\bibinfo {author} {\bibfnamefont {L.~M.}\ \bibnamefont
  {Da~Silva}}, \bibinfo {author} {\bibfnamefont {R.}~\bibnamefont {Cesar}},
  \bibinfo {author} {\bibfnamefont {C.~M.}\ \bibnamefont {Moreira}}, \bibinfo
  {author} {\bibfnamefont {J.~H.}\ \bibnamefont {Santos}}, \bibinfo {author}
  {\bibfnamefont {L.~G.}\ \bibnamefont {De~Souza}}, \bibinfo {author}
  {\bibfnamefont {B.~M.}\ \bibnamefont {Pires}}, \bibinfo {author}
  {\bibfnamefont {R.}~\bibnamefont {Vicentini}}, \bibinfo {author}
  {\bibfnamefont {W.}~\bibnamefont {Nunes}},\ and\ \bibinfo {author}
  {\bibfnamefont {H.}~\bibnamefont {Zanin}},\ }\href@noop {} {\bibfield
  {journal} {\bibinfo  {journal} {Energy storage materials}\ }\textbf {\bibinfo
  {volume} {27}},\ \bibinfo {pages} {555} (\bibinfo {year} {2020})}\BibitemShut
  {NoStop}%
\bibitem [{\citenamefont {Li}\ \emph {et~al.}(2018)\citenamefont {Li},
  \citenamefont {Shao}, \citenamefont {Kim}, \citenamefont {Yao}, \citenamefont
  {Wang}, \citenamefont {Miao}, \citenamefont {Zheng}, \citenamefont {Sun},
  \citenamefont {Zhang},\ and\ \citenamefont {Braun}}]{li2018high}%
  \BibitemOpen
  \bibfield  {author} {\bibinfo {author} {\bibfnamefont {X.}~\bibnamefont
  {Li}}, \bibinfo {author} {\bibfnamefont {J.}~\bibnamefont {Shao}}, \bibinfo
  {author} {\bibfnamefont {S.-K.}\ \bibnamefont {Kim}}, \bibinfo {author}
  {\bibfnamefont {C.}~\bibnamefont {Yao}}, \bibinfo {author} {\bibfnamefont
  {J.}~\bibnamefont {Wang}}, \bibinfo {author} {\bibfnamefont {Y.-R.}\
  \bibnamefont {Miao}}, \bibinfo {author} {\bibfnamefont {Q.}~\bibnamefont
  {Zheng}}, \bibinfo {author} {\bibfnamefont {P.}~\bibnamefont {Sun}}, \bibinfo
  {author} {\bibfnamefont {R.}~\bibnamefont {Zhang}},\ and\ \bibinfo {author}
  {\bibfnamefont {P.~V.}\ \bibnamefont {Braun}},\ }\href@noop {} {\bibfield
  {journal} {\bibinfo  {journal} {Nature communications}\ }\textbf {\bibinfo
  {volume} {9}},\ \bibinfo {pages} {2578} (\bibinfo {year} {2018})}\BibitemShut
  {NoStop}%
\bibitem [{\citenamefont {Budkov}\ and\ \citenamefont
  {Kolesnikov}(2022)}]{budkov2022modified}%
  \BibitemOpen
  \bibfield  {author} {\bibinfo {author} {\bibfnamefont {Y.~A.}\ \bibnamefont
  {Budkov}}\ and\ \bibinfo {author} {\bibfnamefont {A.~L.}\ \bibnamefont
  {Kolesnikov}},\ }\href@noop {} {\bibfield  {journal} {\bibinfo  {journal}
  {Journal of Statistical Mechanics: Theory and Experiment}\ }\textbf {\bibinfo
  {volume} {2022}},\ \bibinfo {pages} {053205} (\bibinfo {year}
  {2022})}\BibitemShut {NoStop}%
\bibitem [{\citenamefont {Budkov}\ and\ \citenamefont
  {Kalikin}(2023)}]{budkov2023macroscopic}%
  \BibitemOpen
  \bibfield  {author} {\bibinfo {author} {\bibfnamefont {Y.~A.}\ \bibnamefont
  {Budkov}}\ and\ \bibinfo {author} {\bibfnamefont {N.~N.}\ \bibnamefont
  {Kalikin}},\ }\href@noop {} {\bibfield  {journal} {\bibinfo  {journal}
  {Physical Review E}\ }\textbf {\bibinfo {volume} {107}},\ \bibinfo {pages}
  {024503} (\bibinfo {year} {2023})}\BibitemShut {NoStop}%
\bibitem [{\citenamefont {Brandyshev}\ and\ \citenamefont
  {Budkov}(2023)}]{brandyshev2023noether}%
  \BibitemOpen
  \bibfield  {author} {\bibinfo {author} {\bibfnamefont {P.~E.}\ \bibnamefont
  {Brandyshev}}\ and\ \bibinfo {author} {\bibfnamefont {Y.~A.}\ \bibnamefont
  {Budkov}},\ }\href@noop {} {\bibfield  {journal} {\bibinfo  {journal} {The
  Journal of chemical physics}\ }\textbf {\bibinfo {volume} {158}},\ \bibinfo
  {pages} {174114} (\bibinfo {year} {2023})}\BibitemShut {NoStop}%
\bibitem [{\citenamefont {Gongadze}\ \emph {et~al.}(2014)\citenamefont
  {Gongadze}, \citenamefont {Velikonja}, \citenamefont {Perutkova},
  \citenamefont {Kramar}, \citenamefont {Ma{\v{c}}ek-Lebar}, \citenamefont
  {Kralj-Igli{\v{c}}},\ and\ \citenamefont {Igli{\v{c}}}}]{gongadze2014ions}%
  \BibitemOpen
  \bibfield  {author} {\bibinfo {author} {\bibfnamefont {E.}~\bibnamefont
  {Gongadze}}, \bibinfo {author} {\bibfnamefont {A.}~\bibnamefont {Velikonja}},
  \bibinfo {author} {\bibfnamefont {{\v{S}}.}~\bibnamefont {Perutkova}},
  \bibinfo {author} {\bibfnamefont {P.}~\bibnamefont {Kramar}}, \bibinfo
  {author} {\bibfnamefont {A.}~\bibnamefont {Ma{\v{c}}ek-Lebar}}, \bibinfo
  {author} {\bibfnamefont {V.}~\bibnamefont {Kralj-Igli{\v{c}}}},\ and\
  \bibinfo {author} {\bibfnamefont {A.}~\bibnamefont {Igli{\v{c}}}},\
  }\href@noop {} {\bibfield  {journal} {\bibinfo  {journal} {Electrochimica
  Acta}\ }\textbf {\bibinfo {volume} {126}},\ \bibinfo {pages} {42} (\bibinfo
  {year} {2014})}\BibitemShut {NoStop}%
\bibitem [{\citenamefont {Marcovitz}, \citenamefont {Naftaly},\ and\
  \citenamefont {Levy}(2015)}]{marcovitz2015water}%
  \BibitemOpen
  \bibfield  {author} {\bibinfo {author} {\bibfnamefont {A.}~\bibnamefont
  {Marcovitz}}, \bibinfo {author} {\bibfnamefont {A.}~\bibnamefont {Naftaly}},\
  and\ \bibinfo {author} {\bibfnamefont {Y.}~\bibnamefont {Levy}},\ }\href@noop
  {} {\bibfield  {journal} {\bibinfo  {journal} {The Journal of Chemical
  Physics}\ }\textbf {\bibinfo {volume} {142}},\ \bibinfo {pages} {02B614\_1}
  (\bibinfo {year} {2015})}\BibitemShut {NoStop}%
\bibitem [{\citenamefont {Maggs}\ and\ \citenamefont
  {Podgornik}(2016)}]{maggs2016general}%
  \BibitemOpen
  \bibfield  {author} {\bibinfo {author} {\bibfnamefont {A.}~\bibnamefont
  {Maggs}}\ and\ \bibinfo {author} {\bibfnamefont {R.}~\bibnamefont
  {Podgornik}},\ }\href@noop {} {\bibfield  {journal} {\bibinfo  {journal}
  {Soft matter}\ }\textbf {\bibinfo {volume} {12}},\ \bibinfo {pages} {1219}
  (\bibinfo {year} {2016})}\BibitemShut {NoStop}%
\bibitem [{\citenamefont {Bazant}, \citenamefont {Storey},\ and\ \citenamefont
  {Kornyshev}(2011)}]{bazant2011double}%
  \BibitemOpen
  \bibfield  {author} {\bibinfo {author} {\bibfnamefont {M.~Z.}\ \bibnamefont
  {Bazant}}, \bibinfo {author} {\bibfnamefont {B.~D.}\ \bibnamefont {Storey}},\
  and\ \bibinfo {author} {\bibfnamefont {A.~A.}\ \bibnamefont {Kornyshev}},\
  }\href@noop {} {\bibfield  {journal} {\bibinfo  {journal} {Physical review
  letters}\ }\textbf {\bibinfo {volume} {106}},\ \bibinfo {pages} {046102}
  (\bibinfo {year} {2011})}\BibitemShut {NoStop}%
\bibitem [{\citenamefont {Blossey}, \citenamefont {Maggs},\ and\ \citenamefont
  {Podgornik}(2017)}]{blossey2017structural}%
  \BibitemOpen
  \bibfield  {author} {\bibinfo {author} {\bibfnamefont {R.}~\bibnamefont
  {Blossey}}, \bibinfo {author} {\bibfnamefont {A.}~\bibnamefont {Maggs}},\
  and\ \bibinfo {author} {\bibfnamefont {R.}~\bibnamefont {Podgornik}},\
  }\href@noop {} {\bibfield  {journal} {\bibinfo  {journal} {Physical Review
  E}\ }\textbf {\bibinfo {volume} {95}},\ \bibinfo {pages} {060602} (\bibinfo
  {year} {2017})}\BibitemShut {NoStop}%
\bibitem [{\citenamefont {Cahn}(1965)}]{cahn1965phase}%
  \BibitemOpen
  \bibfield  {author} {\bibinfo {author} {\bibfnamefont {J.~W.}\ \bibnamefont
  {Cahn}},\ }\href@noop {} {\bibfield  {journal} {\bibinfo  {journal} {The
  Journal of chemical physics}\ }\textbf {\bibinfo {volume} {42}},\ \bibinfo
  {pages} {93} (\bibinfo {year} {1965})}\BibitemShut {NoStop}%
\bibitem [{\citenamefont {Borukhov}, \citenamefont {Andelman},\ and\
  \citenamefont {Orland}(1997)}]{borukhov1997steric}%
  \BibitemOpen
  \bibfield  {author} {\bibinfo {author} {\bibfnamefont {I.}~\bibnamefont
  {Borukhov}}, \bibinfo {author} {\bibfnamefont {D.}~\bibnamefont {Andelman}},\
  and\ \bibinfo {author} {\bibfnamefont {H.}~\bibnamefont {Orland}},\
  }\href@noop {} {\bibfield  {journal} {\bibinfo  {journal} {Physical review
  letters}\ }\textbf {\bibinfo {volume} {79}},\ \bibinfo {pages} {435}
  (\bibinfo {year} {1997})}\BibitemShut {NoStop}%
\bibitem [{\citenamefont {Kornyshev}(2007)}]{kornyshev2007double}%
  \BibitemOpen
  \bibfield  {author} {\bibinfo {author} {\bibfnamefont {A.~A.}\ \bibnamefont
  {Kornyshev}},\ }\href@noop {} {\bibfield  {journal} {\bibinfo  {journal} {The
  Journal of Physical Chemistry B}\ }\textbf {\bibinfo {volume} {111}},\
  \bibinfo {pages} {5545} (\bibinfo {year} {2007})}\BibitemShut {NoStop}%
\bibitem [{\citenamefont {Kralj-Igli{\v{c}}}\ and\ \citenamefont
  {Igli{\v{c}}}(1996)}]{kralj1996simple}%
  \BibitemOpen
  \bibfield  {author} {\bibinfo {author} {\bibfnamefont {V.}~\bibnamefont
  {Kralj-Igli{\v{c}}}}\ and\ \bibinfo {author} {\bibfnamefont {A.}~\bibnamefont
  {Igli{\v{c}}}},\ }\href@noop {} {\bibfield  {journal} {\bibinfo  {journal}
  {Journal de Physique II}\ }\textbf {\bibinfo {volume} {6}},\ \bibinfo {pages}
  {477} (\bibinfo {year} {1996})}\BibitemShut {NoStop}%
\bibitem [{\citenamefont {Budkov}\ \emph {et~al.}(2018)\citenamefont {Budkov},
  \citenamefont {Kolesnikov}, \citenamefont {Goodwin}, \citenamefont
  {Kiselev},\ and\ \citenamefont {Kornyshev}}]{budkov2018theory}%
  \BibitemOpen
  \bibfield  {author} {\bibinfo {author} {\bibfnamefont {Y.~A.}\ \bibnamefont
  {Budkov}}, \bibinfo {author} {\bibfnamefont {A.~L.}\ \bibnamefont
  {Kolesnikov}}, \bibinfo {author} {\bibfnamefont {Z.~A.}\ \bibnamefont
  {Goodwin}}, \bibinfo {author} {\bibfnamefont {M.~G.}\ \bibnamefont
  {Kiselev}},\ and\ \bibinfo {author} {\bibfnamefont {A.~A.}\ \bibnamefont
  {Kornyshev}},\ }\href@noop {} {\bibfield  {journal} {\bibinfo  {journal}
  {Electrochimica Acta}\ }\textbf {\bibinfo {volume} {284}},\ \bibinfo {pages}
  {346} (\bibinfo {year} {2018})}\BibitemShut {NoStop}%
\bibitem [{\citenamefont {Igli{\v{c}}}, \citenamefont {Gongadze},\ and\
  \citenamefont {Bohinc}(2010)}]{iglivc2010excluded}%
  \BibitemOpen
  \bibfield  {author} {\bibinfo {author} {\bibfnamefont {A.}~\bibnamefont
  {Igli{\v{c}}}}, \bibinfo {author} {\bibfnamefont {E.}~\bibnamefont
  {Gongadze}},\ and\ \bibinfo {author} {\bibfnamefont {K.}~\bibnamefont
  {Bohinc}},\ }\href@noop {} {\bibfield  {journal} {\bibinfo  {journal}
  {Bioelectrochemistry}\ }\textbf {\bibinfo {volume} {79}},\ \bibinfo {pages}
  {223} (\bibinfo {year} {2010})}\BibitemShut {NoStop}%
\bibitem [{\citenamefont {Budkov}, \citenamefont {Kolesnikov},\ and\
  \citenamefont {Kiselev}(2016)}]{budkov2016theory}%
  \BibitemOpen
  \bibfield  {author} {\bibinfo {author} {\bibfnamefont {Y.~A.}\ \bibnamefont
  {Budkov}}, \bibinfo {author} {\bibfnamefont {A.}~\bibnamefont {Kolesnikov}},\
  and\ \bibinfo {author} {\bibfnamefont {M.}~\bibnamefont {Kiselev}},\
  }\href@noop {} {\bibfield  {journal} {\bibinfo  {journal} {The Journal of
  chemical physics}\ }\textbf {\bibinfo {volume} {144}},\ \bibinfo {pages}
  {184703} (\bibinfo {year} {2016})}\BibitemShut {NoStop}%
\bibitem [{\citenamefont {Budkov}, \citenamefont {Kolesnikov},\ and\
  \citenamefont {Kiselev}(2015)}]{budkov2015modified}%
  \BibitemOpen
  \bibfield  {author} {\bibinfo {author} {\bibfnamefont {Y.~A.}\ \bibnamefont
  {Budkov}}, \bibinfo {author} {\bibfnamefont {A.}~\bibnamefont {Kolesnikov}},\
  and\ \bibinfo {author} {\bibfnamefont {M.}~\bibnamefont {Kiselev}},\
  }\href@noop {} {\bibfield  {journal} {\bibinfo  {journal} {Europhysics
  Letters}\ }\textbf {\bibinfo {volume} {111}},\ \bibinfo {pages} {28002}
  (\bibinfo {year} {2015})}\BibitemShut {NoStop}%
\bibitem [{\citenamefont {Nakayama}\ and\ \citenamefont
  {Andelman}(2015)}]{nakayama2015differential}%
  \BibitemOpen
  \bibfield  {author} {\bibinfo {author} {\bibfnamefont {Y.}~\bibnamefont
  {Nakayama}}\ and\ \bibinfo {author} {\bibfnamefont {D.}~\bibnamefont
  {Andelman}},\ }\href@noop {} {\bibfield  {journal} {\bibinfo  {journal} {The
  Journal of chemical physics}\ }\textbf {\bibinfo {volume} {142}},\ \bibinfo
  {pages} {044706} (\bibinfo {year} {2015})}\BibitemShut {NoStop}%
\bibitem [{\citenamefont {Virtanen}\ \emph {et~al.}(2020)\citenamefont
  {Virtanen}, \citenamefont {Gommers}, \citenamefont {Oliphant}, \citenamefont
  {Haberland}, \citenamefont {Reddy}, \citenamefont {Cournapeau}, \citenamefont
  {Burovski}, \citenamefont {Peterson}, \citenamefont {Weckesser},
  \citenamefont {Bright} \emph {et~al.}}]{virtanen2020scipy}%
  \BibitemOpen
  \bibfield  {author} {\bibinfo {author} {\bibfnamefont {P.}~\bibnamefont
  {Virtanen}}, \bibinfo {author} {\bibfnamefont {R.}~\bibnamefont {Gommers}},
  \bibinfo {author} {\bibfnamefont {T.~E.}\ \bibnamefont {Oliphant}}, \bibinfo
  {author} {\bibfnamefont {M.}~\bibnamefont {Haberland}}, \bibinfo {author}
  {\bibfnamefont {T.}~\bibnamefont {Reddy}}, \bibinfo {author} {\bibfnamefont
  {D.}~\bibnamefont {Cournapeau}}, \bibinfo {author} {\bibfnamefont
  {E.}~\bibnamefont {Burovski}}, \bibinfo {author} {\bibfnamefont
  {P.}~\bibnamefont {Peterson}}, \bibinfo {author} {\bibfnamefont
  {W.}~\bibnamefont {Weckesser}}, \bibinfo {author} {\bibfnamefont
  {J.}~\bibnamefont {Bright}}, \emph {et~al.},\ }\href@noop {} {\bibfield
  {journal} {\bibinfo  {journal} {Nature methods}\ }\textbf {\bibinfo {volume}
  {17}},\ \bibinfo {pages} {261} (\bibinfo {year} {2020})}\BibitemShut
  {NoStop}%
\bibitem [{\citenamefont {Derjaguin}, \citenamefont {Churaev},\ and\
  \citenamefont {Muller}(1987)}]{derjaguin1987derjaguin}%
  \BibitemOpen
  \bibfield  {author} {\bibinfo {author} {\bibfnamefont {B.}~\bibnamefont
  {Derjaguin}}, \bibinfo {author} {\bibfnamefont {N.}~\bibnamefont {Churaev}},\
  and\ \bibinfo {author} {\bibfnamefont {V.}~\bibnamefont {Muller}},\ }in\
  \href@noop {} {\emph {\bibinfo {booktitle} {Surface Forces}}}\ (\bibinfo
  {publisher} {Springer},\ \bibinfo {year} {1987})\ pp.\ \bibinfo {pages}
  {293--310}\BibitemShut {NoStop}%
\bibitem [{\citenamefont
  {Israelachvili}(2011)}]{israelachvili2011intermolecular}%
  \BibitemOpen
  \bibfield  {author} {\bibinfo {author} {\bibfnamefont {J.~N.}\ \bibnamefont
  {Israelachvili}},\ }\href@noop {} {\emph {\bibinfo {title} {Intermolecular
  and surface forces}}}\ (\bibinfo  {publisher} {Academic press},\ \bibinfo
  {year} {2011})\BibitemShut {NoStop}%
\bibitem [{\citenamefont {Ciach}(2018)}]{ciach2018simple}%
  \BibitemOpen
  \bibfield  {author} {\bibinfo {author} {\bibfnamefont {A.}~\bibnamefont
  {Ciach}},\ }\href@noop {} {\bibfield  {journal} {\bibinfo  {journal} {Journal
  of Molecular Liquids}\ }\textbf {\bibinfo {volume} {270}},\ \bibinfo {pages}
  {138} (\bibinfo {year} {2018})}\BibitemShut {NoStop}%
\bibitem [{\citenamefont {Zelko}\ \emph {et~al.}(2010)\citenamefont {Zelko},
  \citenamefont {Igli{\v{c}}}, \citenamefont {Kralj-Igli{\v{c}}},\ and\
  \citenamefont {Kumar}}]{zelko2010effects}%
  \BibitemOpen
  \bibfield  {author} {\bibinfo {author} {\bibfnamefont {J.}~\bibnamefont
  {Zelko}}, \bibinfo {author} {\bibfnamefont {A.}~\bibnamefont {Igli{\v{c}}}},
  \bibinfo {author} {\bibfnamefont {V.}~\bibnamefont {Kralj-Igli{\v{c}}}},\
  and\ \bibinfo {author} {\bibfnamefont {P.}~\bibnamefont {Kumar}},\
  }\href@noop {} {\bibfield  {journal} {\bibinfo  {journal} {The Journal of
  chemical physics}\ }\textbf {\bibinfo {volume} {133}} (\bibinfo {year}
  {2010})}\BibitemShut {NoStop}%
\bibitem [{\citenamefont {Barrat}\ and\ \citenamefont
  {Hansen}(2003)}]{barrat2003basic}%
  \BibitemOpen
  \bibfield  {author} {\bibinfo {author} {\bibfnamefont {J.-L.}\ \bibnamefont
  {Barrat}}\ and\ \bibinfo {author} {\bibfnamefont {J.-P.}\ \bibnamefont
  {Hansen}},\ }\href@noop {} {\emph {\bibinfo {title} {Basic concepts for
  simple and complex liquids}}}\ (\bibinfo  {publisher} {Cambridge University
  Press},\ \bibinfo {year} {2003})\BibitemShut {NoStop}%
\bibitem [{\citenamefont {Taboada-Serrano}, \citenamefont {Yiacoumi},\ and\
  \citenamefont {Tsouris}(2006)}]{taboada2006electrostatic}%
  \BibitemOpen
  \bibfield  {author} {\bibinfo {author} {\bibfnamefont {P.}~\bibnamefont
  {Taboada-Serrano}}, \bibinfo {author} {\bibfnamefont {S.}~\bibnamefont
  {Yiacoumi}},\ and\ \bibinfo {author} {\bibfnamefont {C.}~\bibnamefont
  {Tsouris}},\ }\href@noop {} {\bibfield  {journal} {\bibinfo  {journal} {The
  Journal of chemical physics}\ }\textbf {\bibinfo {volume} {125}},\ \bibinfo
  {pages} {054716} (\bibinfo {year} {2006})}\BibitemShut {NoStop}%
\bibitem [{\citenamefont {Kewalramani}\ \emph {et~al.}(2016)\citenamefont
  {Kewalramani}, \citenamefont {Guerrero-Garc{\'\i}a}, \citenamefont {Moreau},
  \citenamefont {Zwanikken}, \citenamefont {Mirkin}, \citenamefont {Olvera
  de~la Cruz},\ and\ \citenamefont {Bedzyk}}]{kewalramani2016electrolyte}%
  \BibitemOpen
  \bibfield  {author} {\bibinfo {author} {\bibfnamefont {S.}~\bibnamefont
  {Kewalramani}}, \bibinfo {author} {\bibfnamefont {G.~I.}\ \bibnamefont
  {Guerrero-Garc{\'\i}a}}, \bibinfo {author} {\bibfnamefont {L.~M.}\
  \bibnamefont {Moreau}}, \bibinfo {author} {\bibfnamefont {J.~W.}\
  \bibnamefont {Zwanikken}}, \bibinfo {author} {\bibfnamefont {C.~A.}\
  \bibnamefont {Mirkin}}, \bibinfo {author} {\bibfnamefont {M.}~\bibnamefont
  {Olvera de~la Cruz}},\ and\ \bibinfo {author} {\bibfnamefont {M.~J.}\
  \bibnamefont {Bedzyk}},\ }\href@noop {} {\bibfield  {journal} {\bibinfo
  {journal} {ACS central science}\ }\textbf {\bibinfo {volume} {2}},\ \bibinfo
  {pages} {219} (\bibinfo {year} {2016})}\BibitemShut {NoStop}%
\bibitem [{\citenamefont {Gongadze}\ and\ \citenamefont
  {Igli{\v{c}}}(2013)}]{gongadze2013excluded}%
  \BibitemOpen
  \bibfield  {author} {\bibinfo {author} {\bibfnamefont {E.}~\bibnamefont
  {Gongadze}}\ and\ \bibinfo {author} {\bibfnamefont {A.}~\bibnamefont
  {Igli{\v{c}}}},\ }\href@noop {} {\bibfield  {journal} {\bibinfo  {journal}
  {Gen Phys Biophys}\ }\textbf {\bibinfo {volume} {21}},\ \bibinfo {pages}
  {143} (\bibinfo {year} {2013})}\BibitemShut {NoStop}%
\bibitem [{\citenamefont {Gongadze}\ and\ \citenamefont
  {Igli{\v{c}}}(2015)}]{gongadze2015asymmetric}%
  \BibitemOpen
  \bibfield  {author} {\bibinfo {author} {\bibfnamefont {E.}~\bibnamefont
  {Gongadze}}\ and\ \bibinfo {author} {\bibfnamefont {A.}~\bibnamefont
  {Igli{\v{c}}}},\ }\href@noop {} {\bibfield  {journal} {\bibinfo  {journal}
  {Electrochimica Acta}\ }\textbf {\bibinfo {volume} {178}},\ \bibinfo {pages}
  {541} (\bibinfo {year} {2015})}\BibitemShut {NoStop}%
\end{thebibliography}%

\end{document}